\documentclass[twocolumn,times]{aastex63}
\usepackage{xspace}
\usepackage{comment}
\usepackage{amsmath}
\usepackage{xcolor}
\usepackage{booktabs}
\usepackage{multirow}

\newcommand{\chandra}{\textit{Chandra}\xspace}
\newcommand{\bootes}{Bo\"otes\xspace}
\newcommand{\xb}{XBo\"otes\xspace}
\newcommand{\lognlogs}{LogN-LogS\xspace}
\newcommand{\dnds}{$dN/dS$\xspace}
\newcommand{\fluxcgs}{erg cm$^{-2}$ s$^{-1}$\xspace}
\newcommand{\lumcgs}{erg s$^{-1}$\xspace}
\newcommand{\beforexb}{6838\xspace}
\newcommand{\xbmissed}{447\xspace}
\newcommand{\xbtba}{53\xspace}
\newcommand{\xbbelowthresh}{394\xspace}
\newcommand{\howmany}{6891\xspace}
\newcommand{\softcxb}{$65.0 \pm 12.8\%$\xspace}
\newcommand{\hardcxb}{$81.0 \pm 11.5\%$\xspace}
\newcommand{\panycut}{0.29\xspace}
\newcommand{\howmanycols}{102\xspace}

\submitjournal{ApJ}

\shorttitle{The CDWFS Survey}
\shortauthors{Masini et al.}

\begin{document}

\title{The \chandra Deep Wide-Field Survey: A New \chandra Legacy Survey in the \bootes Field \\ I. X-ray Point Source Catalog, Number Counts and Multi-Wavelength Counterparts}

\correspondingauthor{Alberto Masini}
\email{amasini@sissa.it}

\author[0000-0002-7100-9366]{Alberto Masini}
\affil{SISSA, Via Bonomea 265, 34151 Trieste, Italy}
\affil{Dartmouth College, 6127 Wilder Laboratory, Hanover, NH, 03755 USA}

\author[0000-0003-1468-9526]{Ryan C. Hickox}
\affiliation{Dartmouth College, 6127 Wilder Laboratory, Hanover, NH, 03755 USA}

\author[0000-0003-3574-2963]{Christopher M. Carroll}
\affiliation{Dartmouth College, 6127 Wilder Laboratory, Hanover, NH, 03755 USA}

\author[0000-0003-1908-8463]{James Aird}
\affiliation{Department of Physics \& Astronomy, University of Leicester, University Road, Leicester LE1 7RJ, UK}

\author[0000-0002-5896-6313]{David M. Alexander}
\affiliation{Centre for Extragalactic Astronomy, Department of Physics, Durham University, South Road, Durham, DH1 3LE, UK}

\author[0000-0002-9508-3667]{Roberto J. Assef}
\affiliation{N\'ucleo de Astronom\'ia de la Facultad de Ingenier\'ia y Ciencias, Universidad Diego Portales, Av. Ej\'ercito Libertador 441, Santiago, Chile}

\author[0000-0002-5215-6010]{Richard Bower}
\affiliation{Institute for Computational Cosmology, Department of Physics, University of Durham, South Road, Durham DH1 3LE, UK}

\author[0000-0002-4208-798X]{Mark Brodwin}
\affiliation{University of Missouri, 5110 Rockhill Road, Kansas City, MO 64110, USA}

\author[0000-0002-1207-9137]{Michael J.~I. Brown}
\affiliation{School of Physics and Astronomy, Monash University, Clayton 3800, Victoria, Australia}

\author[0000-0002-3236-2853]{Suchetana Chatterjee}
\affiliation{Department of Physics, Presidency University, Kolkata, 700073, India}

\author[0000-0002-4945-5079]{Chien-Ting J. Chen}
\affiliation{Marshall Space Flight Center, Huntsville, AL 35811, USA}

\author[0000-0002-4928-4003]{Arjun Dey}
\affiliation{National Optical Astronomy Observatory, 950 N. Cherry Avenue, Tucson, AZ 85719, USA}

\author[0000-0001-6788-1701]{Michael A. DiPompeo}
\affiliation{Dartmouth College, 6127 Wilder Laboratory, Hanover, NH, 03755 USA}

\author[0000-0001-6889-8388]{Kenneth J. Duncan}
\affil{SUPA, Institute for Astronomy, Royal Observatory, Blackford Hill, Edinburgh, EH9 3HJ, UK}
\affil{Leiden Observatory, Leiden University, PO Box 9513, NL-2300 RA Leiden, The Netherlands}

\author{Peter R.~M. Eisenhardt}
\affiliation{Jet Propulsion Laboratory, California Institute of Technology, 4800 Oak Grove Drive, M/S 169-327, Pasadena, CA91109, USA}

\author[0000-0002-9478-1682]{William R. Forman}
\affiliation{Center for Astrophysics $\vert$ Harvard \& Smithsonian, 60 Garden St, Cambridge, MA 02138, USA}

\author[0000-0002-0933-8601]{Anthony H. Gonzalez}
\affiliation{Department of Astronomy, University of Florida, 211 Bryant Space Center, Gainesville, FL 32611, USA}

\author{Andrew D. Goulding}
\affiliation{Department of Astrophysical Sciences, Princeton University, Ivy Lane, Princeton, NJ 08544, USA}

\author[0000-0003-4565-8239]{Kevin N. Hainline}
\affiliation{Steward Observatory, University of Arizona, 933 North Cherry Avenue, Tucson, AZ 85721, USA}

\author[0000-0002-1578-6582]{Buell T. Jannuzi}
\affiliation{Steward Observatory, University of Arizona, 933 North Cherry Avenue, Tucson, AZ 85721, USA}

\author{Christine Jones}
\affiliation{Center for Astrophysics $\vert$ Harvard \& Smithsonian, 60 Garden St, Cambridge, MA 02138, USA}

\author[0000-0001-6017-2961]{Christopher S. Kochanek}
\affiliation{Department of Astronomy, Ohio State University, 140 West 18th Avenue, Columbus, OH 43210, USA}

\author[0000-0002-0765-0511]{Ralph Kraft}
\affiliation{Center for Astrophysics $\vert$ Harvard \& Smithsonian, 60 Garden St, Cambridge, MA 02138, USA}

\author[0000-0003-3004-9596]{Kyoung-Soo Lee}
\affiliation{Department of Physics, Purdue University, 525 Northwestern Avenue, West Lafayette, IN 47907, USA}

\author[0000-0002-3031-2326]{Eric D. Miller}
\affiliation{Kavli Institute for Astrophysics and Space Research, Massachusetts Institute of Technology, Cambridge, MA 02139, USA}

\author[0000-0002-3126-6712]{James Mullaney}
\affiliation{Department of Physics and Astronomy, The University of Sheffield, Hounsfield Road, Sheffield S3 7RH, UK}

\author{Adam D. Myers}
\affiliation{Department of Physics \& Astronomy, University of Wyoming, 1000 University Ave., Laramie, WY 82071, USA}

\author[0000-0001-5655-1440]{Andrew Ptak}
\affiliation{NASA/Goddard Spaceflight Center, Mail Code 662, Greenbelt, MD 20771, USA}

\author{Adam Stanford}
\affiliation{Physics Department, University of California, Davis, CA 95616, USA}

\author[0000-0003-2686-9241]{Daniel Stern}
\affiliation{Jet Propulsion Laboratory, California Institute of Technology, 4800 Oak Grove Drive, M/S 169-327, Pasadena, CA 91109, USA}

\author[0000-0001-8121-0234]{Alexey Vikhlinin}
\affiliation{Center for Astrophysics $\vert$ Harvard \& Smithsonian, 60 Garden St, Cambridge, MA 02138, USA}

\author[0000-0002-6047-1010]{David A. Wake}
\affiliation{Department of Physics, University of North Carolina Asheville, One University Heights, Asheville, NC 28804, USA}
\affiliation{Department of Physical Sciences, The Open University, Milton Keynes MK7 6AA, UK}

\author{Stephen S. Murray}
\altaffiliation{Steve Murray passed away on 2015 August 10. This survey would not have been possible without his invaluable contribution.}
\affiliation{Center for Astrophysics $\vert$ Harvard \& Smithsonian, 60 Garden St, Cambridge, MA 02138, USA}

\begin{abstract}

We present a new, ambitious survey performed with the \chandra \textit{X-ray Observatory} of the 9.3 deg$^2$ Bo\"otes field of the NOAO Deep Wide-Field Survey. The wide field probes a statistically representative volume of the Universe at high redshift. The \chandra Deep Wide-Field Survey exploits the excellent sensitivity and angular resolution of \chandra over a wide area, combining 281 observations spanning 15 years, for a total exposure time of 3.4 Ms, and detects \howmany X-ray point sources down to limiting fluxes of $4.7\times10^{-16}$, $1.5\times10^{-16}$, and $9\times10^{-16}$ \fluxcgs, in the $0.5-7$ keV, $0.5-2$ keV, and $2-7$ keV bands, respectively. The robustness and reliability of the detection strategy is validated through extensive, state-of-the-art simulations of the whole field.
Accurate number counts, in good agreement with previous X-ray surveys, are derived thanks to the uniquely large number of point sources detected, which resolve \softcxb of the cosmic X-ray background between $0.5-2$ keV and \hardcxb between $2-7$ keV. Exploiting the wealth of multi-wavelength data available on the field, we assign redshifts to $\sim 94\%$ of the X-ray sources, estimate their obscuration and derive absorption-corrected luminosities.
We provide an electronic catalog containing all the relevant quantities needed for future investigations.

\end{abstract}


\section{Introduction} \label{sec:intro}

It is now widely accepted that the large majority of massive galaxies in the observable Universe host Supermassive Black Holes (SMBHs) in their nuclei, with masses ranging between $10^6$--$10^{10} M_{\odot}$ \citep{kormendyho13}. The evidence is provided by multiple observations, such as tracing the kinematics and dynamics of stars in the central region of the Milky Way \citep[e.g.,][]{ghez08} and in the bulges of nearby galaxies \citep{kormendyrichstone95,kormendy04}, from water megamasers \citep{kuo11}, and the recent direct imaging of the shadow of the SMBH in M87 \citep{eht19}. 
\par Studies of SMBHs have gained increasing importance in the last decades, subsequent to the discovery of scaling relations connecting the masses of SMBHs to properties of their host galaxy bulges, such as bulge luminosity \citep{magorrian98}, mass \citep{haeringrix04}, and stellar velocity dispersion \citep{ferraresemerritt00,gebhardt00}, suggesting a co-evolution of the two. In particular, there must be some connection between the growth of the SMBH (which happens through direct accretion of matter and presumably through mergers), and the mass growth of the environment in which it resides (i.e., as traced by star formation).
\par At any given time, the majority of SMBHs lie dormant, but a small fraction is known to emit a significant amount of light and energy, often outshining the whole host galaxy across the electromagnetic spectrum, resulting in an Active Galactic Nucleus (AGN). AGNs are powered by the release of gravitational energy from matter accreted onto the SMBHs in the nuclei of galaxies. The large energy released by an AGN can effectively influence its surrounding environment, eventually impacting the whole bulge/nuclear region, as well as the star formation and the evolution of the galaxy as a whole \citep[see, e.g.][for a recent review]{harrison17}. Strong observational evidence of AGN feedback \citep[e.g.,][]{fabian12review} has shown that AGNs are responsible for the observed correlations between the mass of inactive SMBHs and host galaxy properties. The similarity of the cosmic star formation and SMBH accretion rate histories, both peaking in the same redshift range \citep[$z\sim 2$, see][for a review]{madaudickinson14}, provides an additional piece of evidence that the two phenomena are linked, although many details are still missing.
\par Indeed, despite AGNs now being much-studied, many compelling questions remain unanswered, given the difficulty of investigating causal links between phenomena that span a large range of spatial and temporal scales (e.g., accretion happens on the $\mu$pc scales, while feedback occurs on kpc scales). Studying and solving these cosmic puzzles is further complicated by differences in the environment, cosmic epoch, luminosity, morphology, color and mass of the galaxies considered, by the obscuration state of the AGN, and so on.
\par Thus, while a generally accepted picture of AGN-galaxy co-evolution has emerged through the years, the exact details of how AGNs impact the overall galaxy population are still elusive. Large, statistically robust samples of AGNs can help address the open questions and potentially lead to new discoveries. However, particular care has to be paid to the different selection methods: it is now well-established that AGNs selected with radio, IR, optical and X-ray observations probe broadly different samples of the underlying parent population of AGNs \citep{hickox09}. Moreover, despite their significant luminosity, the effects of obscuration by dust and gas, and dilution by the light of the stars of the host galaxy, severely hamper any unbiased selection of large, statistical samples \citep{hickoxalexander18}. Usually, the X-ray energy band is considered to be the most reliable in detecting accreting SMBHs, due to high contrast with the host galaxy and the high penetrating power of X-ray radiation \citep{brandtalexander15}. However, even X-ray-selected samples suffer some bias against obscured sources, and heavily rely on multi-wavelength data to derive redshifts (either spectroscopic or photometric), luminosities, and host galaxy properties, such as the stellar mass and star formation rate.
\par Over the last few decades, the large majority of X-ray telescopes have undertaken an ever-growing number of X-ray surveys of the sky, covering large portions of the flux-area plane. A given exposure time can be spent staring for a long time on a small area of the sky \citep[e.g., what has been done for the \chandra deep fields; see ][for a review]{xue17} to detect extremely faint X-ray sources, or spread over a larger area, at the cost of missing the faintest sources to obtain better statistics for bright sources. Ideally, a combination of deep pencil-beam and shallow wide surveys is key to probe the largest portion of the parameter space in terms of AGN luminosity, accretion rate, redshift and obscuration state.
\par The majority of the more extensive \chandra surveys have been updated in the last decade, pushing observations to either deeper flux limits \citep[e.g., accumulating up to 7 Ms of time in the \chandra Deep Field South,][]{xue17,luo17} or wider areas \citep[e.g., extending the \chandra coverage in COSMOS, from the C-COSMOS to the \chandra COSMOS Legacy Survey,][]{elvis09,civano16}. One such survey, the NOAO Deep Wide-Field Survey (NDWFS) \bootes field, has, however, lagged behind in sensitivity.
\par Subsequent to the first extensive \xb survey \citep{murray05,kenter05} more than 15 years ago, the NDWFS \bootes field has been observed multiple times but in a heterogeneous way. So, an additional large program was scheduled to cover the central 6 deg$^2$ of the field with more than 1 Ms of \chandra time. These observations were designed to push the detection limit to significantly fainter sources, and to upgrade the \bootes field to better match other state-of-the-art surveys. In this paper we present the \chandra Deep Wide-Field Survey (CDWFS) which, given its optimal combination of sensitivity and survey area, significantly improves the discovery space in the flux limit-survey area plane (see, e.g., Figure 1 of \citealt{xue17}, and Figure 3 of \citealt{brandtalexander15}). We exploit the full extent of the \chandra data in the NDWFS \bootes field, homogeneously analyzing the plethora of available observations, to produce a new, high-quality dataset which will serve as a base for future scientific discoveries.
\par The paper is structured as follows: we present in detail the data reduction and preparation in \S \ref{sec:data_red}. In \S \ref{sec:sims}, we describe the methodology adopted to build and exploit simulations of the whole CDWFS dataset. Once the probability thresholds are calibrated, the production of the X-ray catalog is presented in \S \ref{sec:sdet}. The large number of X-ray point sources is used to derive accurate number counts in \S \ref{sec:ncts}, while \S \ref{sec:opt} describes the procedure used to assign multiwavelength counterparts and redshifts to our X-ray sources. In the same Section we also derive hardness ratios, column densities, and de-absorbed rest frame luminosities. A brief description of each column in the catalog is given in \S \ref{sec:catalog}. The future potential of CDWFS is summarized in \S \ref{sec:summary}.
\par To derive luminosities, we assume a flat $\Lambda$CDM cosmology with $H_0 = 70$ km s$^{-1}$ Mpc$^{-1}$, and $\Omega_{\rm M}=0.3$.

\section{Data reduction} \label{sec:data_red}

The NDWFS \bootes field \citep{jannuzi&dey99}, centered on coordinates (J2000) of R.A.~$=$~14:32:05.712, DEC. $=$~+34:16:47.496, has been extensively observed by \chandra over more than 15 years of operations. Many additional pointings complemented the two most intensive runs of observations, the \xb survey performed during \chandra's Cycle 3 \citep{murray05, kenter05}, and the recent large program in Cycle 18 (PI R. Hickox).
\par To fully exploit the huge amount of information on the same area of the sky, we collected all 281 \chandra pointings in the \bootes field conducted between 2003 and 2018, all in VFAINT mode. A log showing the first ten observations included in this work can be found in Table \ref{tab:obs} (the complete list of observations is available in electronic format).

\begin{deluxetable}{ccccrcc}

\tablecaption{Description of the first ten \chandra observations considered in this work. The full version can be found in the electronic version of the Journal. \label{tab:obs}}

\tablehead{\colhead{ObsID} & \colhead{R.A.} & \colhead{DEC.} & \colhead{Roll} & \colhead{Exposure} & \colhead{MJD} & \colhead{Cycle} \\ 
\colhead{} & \colhead{(deg)} & \colhead{(deg)} & \colhead{(deg)} & \colhead{(ks)} & \colhead{} & \colhead{} } 

\startdata
3130 &  216.408 &  35.600 &  163.4 &  119.9 & 52380.1 &  3 \\
3482 &  216.407 &  35.596 &  219.9 &  58.5 & 52434.2 &  3 \\
3596 &  219.751 &  35.702 &  141.0 &  4.6 & 52735.0 & 4 \\
3597 &  219.321 &  35.702 &  141.0 &  4.6 & 52734.5 & 4 \\
3598 &  218.891 &  35.702 &  141.0 &  4.7 & 52733.1 & 4 \\
3599 &  218.461 &  35.702 &  141.0 &  4.6 & 52732.2 & 4 \\
3600 &  218.031 &  35.702 &  141.0 &  4.7 & 52730.5 & 4 \\
3601 &  217.601 &  35.702 &  141.0 &  4.5 & 52736.1 & 4 \\
3602 &  217.171 &  35.702 &  141.0 &  4.5 & 52735.0 & 4 \\
3603 &  216.742 &  35.702 &  141.0 &  4.7 & 52725.4 & 4 \\
\enddata


\end{deluxetable}

\subsection{Astrometric Correction}

The 281 \chandra observations were downloaded from the \chandra Data Archive\footnote{\href{https://cda.harvard.edu/chaser/}{https://cda.harvard.edu/chaser/}} and reduced with CIAO 4.11 \citep{fruscione06} and CALDB version 4.8.2.
\par
\begin{figure}
    \centering
    \includegraphics[width=0.47\textwidth]{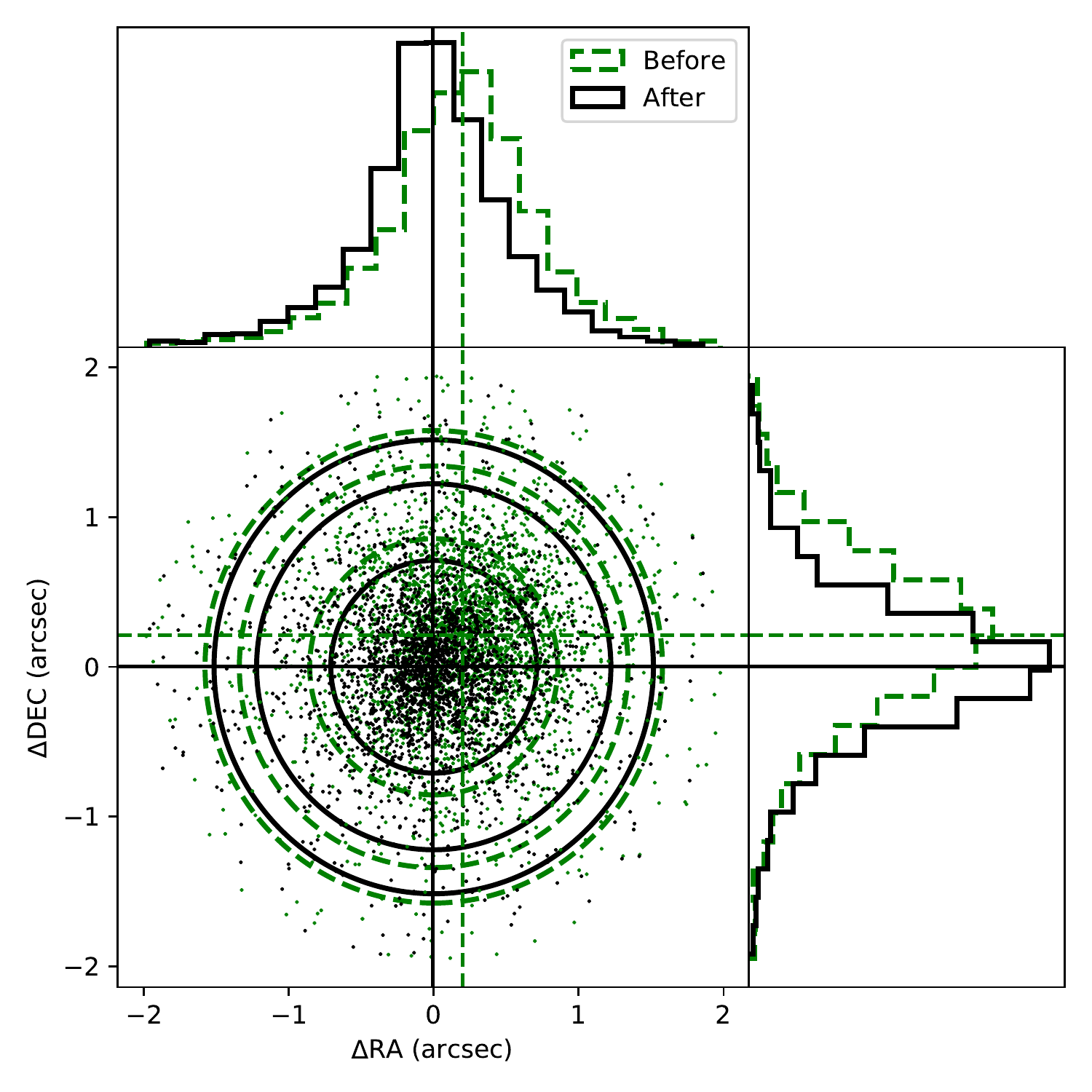}
    \caption{The distribution of the difference in R.A. and DEC. between X-ray and optical sources before (green points and dashed lines) and after (black points and solid lines) the astrometric correction. The three circles represent the radii at which 68\%, 90\% and 95\% of the sources are enclosed, respectively. The alignment centers the distributions on zero, and makes them narrower. As a result, the CDWFS observations are aligned with the NDWFS astrometry.}
    \label{fig:astro_corr}
\end{figure}

 After reprocessing the data with the task \texttt{chandra\_repro} with the flag \texttt{check\_vf\_pha=yes}, the observations were aligned. First, we ran the source detection tool \texttt{wavdetect} on each observation, adopting a permissive threshold (using three scales of ($\sqrt{2}$, 2, 4) pixels and setting the number of allowed spurious sources per scale to be $\texttt{falsesrc}=3$, i.e. allowing for $\sim 10$ spurious sources per field). Then, we matched sources within $6'$ of the aimpoint (where the \chandra point spread function, PSF, is approximately circular) to the catalog of optical counterparts to \xb sources of \citet{brand06}\footnote{This implies that the optical sources are counterparts to previously detected \xb sources with at least four counts.}, using the task \texttt{reproject\_aspect}. This task rotates and translates an observation to minimize the difference between the positions of the X-ray sources and the reference list of optical counterparts. The \texttt{reproject\_aspect} task requires at least three sources in common between the X-ray and optical catalogs to solve for both a rotation and a translation. Six of our observations had fewer than three X-ray/optical sources in common.  For these, we used \texttt{method=trans} instead of the default method. This flag forces \texttt{reproject\_aspect} to use only translation transformations and requires a minimum of one common source. The data were then reprocessed using the new aspect solution files. The results of this calibration step are shown in Figure \ref{fig:astro_corr}, which demonstrates how the calibration
adjusts and narrows the source distribution: 68\% of the data are matched within $0\farcs7$, 90\% within $1\farcs2$ and 95\% within $1\farcs5$. Of course, this approach relies upon the correct identification of optical counterparts to \xb sources; however, the width of the histograms in Figure \ref{fig:astro_corr} is consistent with being largely dominated by X-ray positional uncertainties. We tested that a pure X-ray alignment of the observations (i.e., not relying on optical counterparts, but employing common X-ray sources detected in partially overlapping observations) requires to use very off-axis sources due to the scarcity of overlap for many observations, hampering the accuracy of the astrometric correction. The large exposure time range for partially overlapping observations further complicates the feasibility of such an alignment on the whole 9.3 deg$^2$ area.
\par The described procedure aligned the X-ray observations to the world coordinate system (WCS) of the NDWFS \bootes field, which is the USNO-A2 one. As noted by \citet{cool07}, the USNO-A2 WCS could have a systematic shift with respect to more recent ones. Thus, we decided to register all the data and catalogs that we are going to use throughout this paper to the same astrometric reference, the most recent being the one provided by the \textit{Gaia} mission \citep{gaia16}. Cross-matching USNO-A2 with \textit{Gaia} DR2 \citep{gaiadr2} in the same region of the sky, encompassing the whole \bootes field, we found a systematic shift of RA$_{\rm NDWFS} -$ RA$_{\rm Gaia}$ $= 0\farcs32$ and DEC$_{\rm NDWFS} -$ DEC$_{\rm Gaia}$ $= 0\farcs20$. After checking that the shift is consistent throughout the field, we applied the appropriate astrometric correction to all our mosaics. On the other hand, the \textit{Spitzer} data of the \bootes field that we are going to use in this paper did not show any coordinate shift with respect to \textit{Gaia}'s astrometry. We refer the interested reader to Appendix \ref{asec:astrometry} for a detailed discussion of the astrometry and coordinate registration.

\subsection{Flare Filtering}

We cleaned our observations of time intervals with high-background using the tool \texttt{dmextract} to create $0.5-7$ keV band light curves and \texttt{lc\_sigma\_clip} to create a good time interval (GTI) file after applying $3\sigma$ clipping to the light curve. We then used \texttt{dmcopy} to filter our event files using the GTI files.
\par This procedure resulted in a minimal time loss of 31 ks over $3.4$ Ms of data (less than 1\%). We also note that, as reported by \citet{murray05}, six \xb observations (i.e., ObsIDs 3601, 3607, 3617, 3625, 3641, 3657) had a higher background compared to the others. These observations were not discarded, consistently with the procedure of \citet{murray05}. In addition to these observations, the short (6.1 ks) ObsID 17423 suffered from flaring but was not discarded.
The inclusion of these seven observations is expected to have a negligible impact on the detection of sources given their relatively short exposure.

\subsection{Exposure Map Creation}

Exposure maps were created with the \texttt{fluximage} task. In this work, we adopt the following \chandra bands: full or broad (F; $0.5-7.0$ keV), soft (S; $0.5-2.0$ keV), and hard (H; $2.0-7.0$ keV). When run in its default mode, \texttt{fluximage} produces vignetting-corrected exposure maps in units of ${\rm cm}^{2}\,{\rm s}\,{\rm count}/{\rm photon}$, while if the tool is run with the parameter \texttt{units = time}, the effective area of the \chandra mirrors is ignored. To obtain vignetting-corrected exposure maps with units of seconds, we then produced effective area maps running \texttt{fluximage} with \texttt{units = area}~(i.e., units of ${\rm cm}^{2}\,{\rm count}/{\rm photon}$), and then divided each default exposure map by the maximum value of its corresponding effective area map.
\par
\begin{figure}
    \centering
    \includegraphics[width=0.5\textwidth]{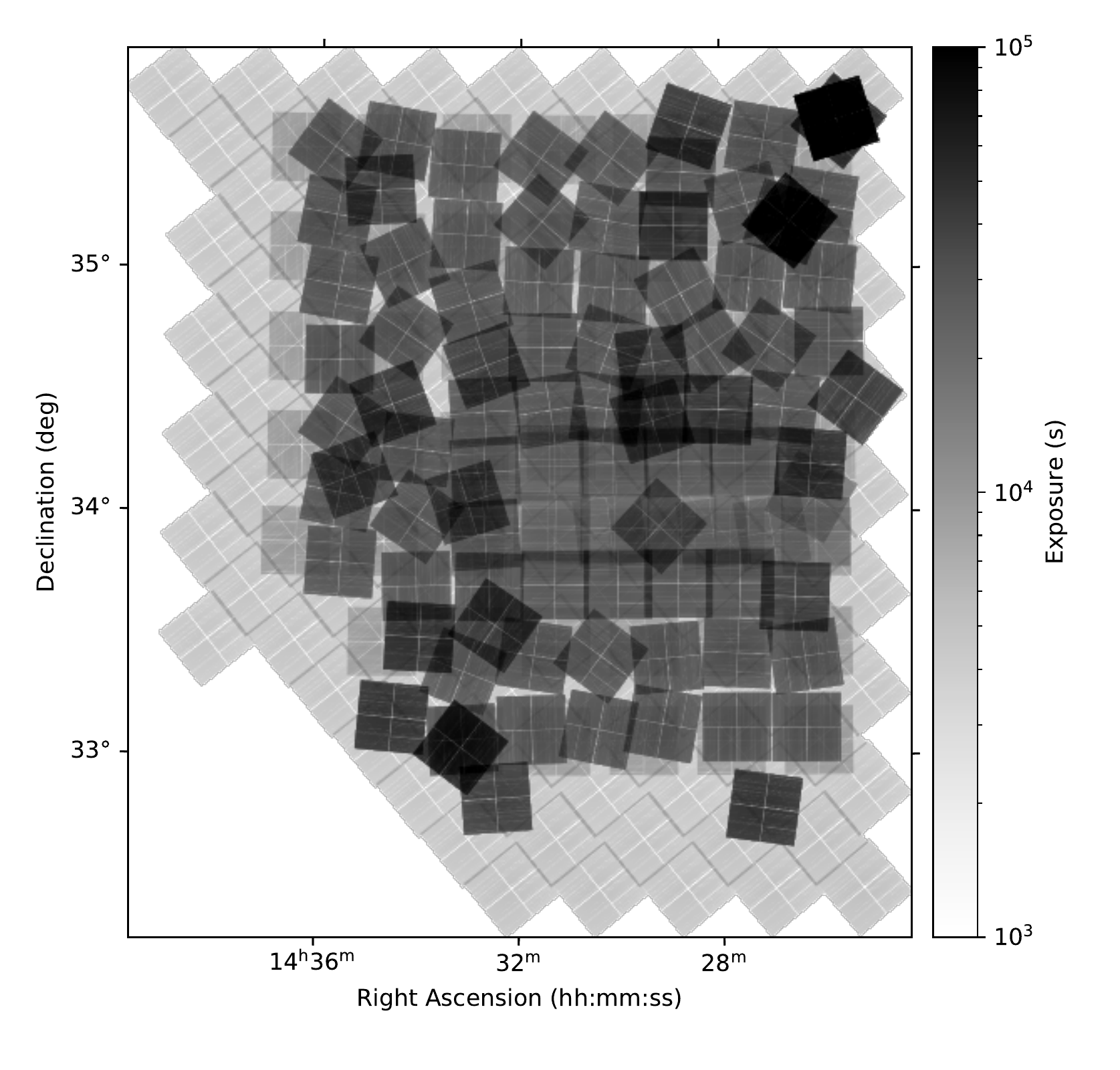}
    \caption{The exposure map of the whole \bootes field shows the 281 overlapping observations considered in this work, as well as the large dynamic range in exposure involved. The external part of the field shows the observations of the \xb survey at broadly $\sim 5$ ks of exposure (light grey), while the \chandra Cycle 18 large program (darker grey) pushes the central 6 deg$^2$ of the field to $\sim 30$ ks of depth. Other deep pointed observations (black) are also included.}
    \label{fig:expomap}
\end{figure}

\begin{figure*}
    \centering
    \includegraphics[width=\textwidth]{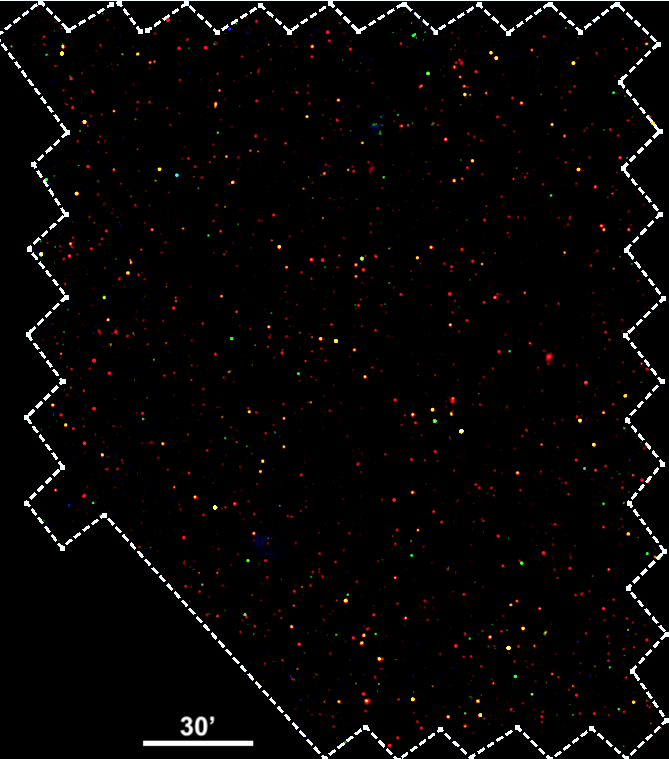}
    \caption{False color image of the whole \bootes field as observed by \chandra, for illustration purpose only. We mapped the $0.5-2$ keV band as red, $2-4.5$ keV as green, and $4.5-7$ keV as blue. Due to the large size of the field, we smoothed the whole image with a Gaussian filter of 20 pixels (i.e. 39\farcs36) radius, and both the scale and color bar were adjusted to enhance the contrast and color of the sources. Due to the large dynamic range in exposure and heavy smoothing, many sources in the deeper part of the field and having a smaller PSF are not clearly visible. The dashed white line delimits the field, while the solid white segment labels $30\arcmin$. North is up and East on the left.}
    \label{fig:tricolor}
\end{figure*}

 Since the exposure map is theoretically monochromatic, it is computed at a given effective energy. The effective energies adopted in this work are 2.3 keV, 1.5 keV, and 3.8 keV for the F, S, and H bands, respectively. The F-band exposure map of the whole field is shown in Figure \ref{fig:expomap}.
All of the mosaics used in this work were created with the CIAO task \texttt{reproject\_image}, and spatially rebinned by a factor of four to speed up the computation time. This means that each pixel of the $4\times4$ rebinned mosaics has a scale of 1\farcs968. A false color image of the whole field
is shown in Figure \ref{fig:tricolor}.

\subsection{Background Map Creation}

Precise background maps are crucial to having a reliable detection strategy and to building accurate simulations. The quiescent (i.e., non-flaring) \chandra background can be explained, to first order, by the sum of an instrumental and an astrophysical component. The instrumental background is due to particles mimicking the signal of X-ray photons, and to non-astrophysical X-ray photons, such as the ones produced by the interactions of particles with the spacecraft structure. The astrophysical background is the contribution from unresolved X-ray sources, mostly AGNs, which are fainter than the flux limit of each observation (i.e. the so called Cosmic X-ray Background, CXB). In addition, there is also diffuse emission of mainly soft X-rays from various local components, such as the Local Bubble and solar wind charge exchange mechanism \citep{markevitch03, slavin13}. Although the \chandra background is generally very low, the large dynamic range in exposure times over our field requires a careful disentangling of the two components for each observation: indeed, while the instrumental background is un-vignetted, the astrophysical component is vignetted, band-dependent (the diffuse soft emission has a minor contribution in the hard band), and exposure-dependent (i.e. the fraction of unresolved sources, hence the CXB contribution, becomes smaller in deeper exposures).
\par
\begin{figure}
    \centering
    \includegraphics[width=0.47\textwidth]{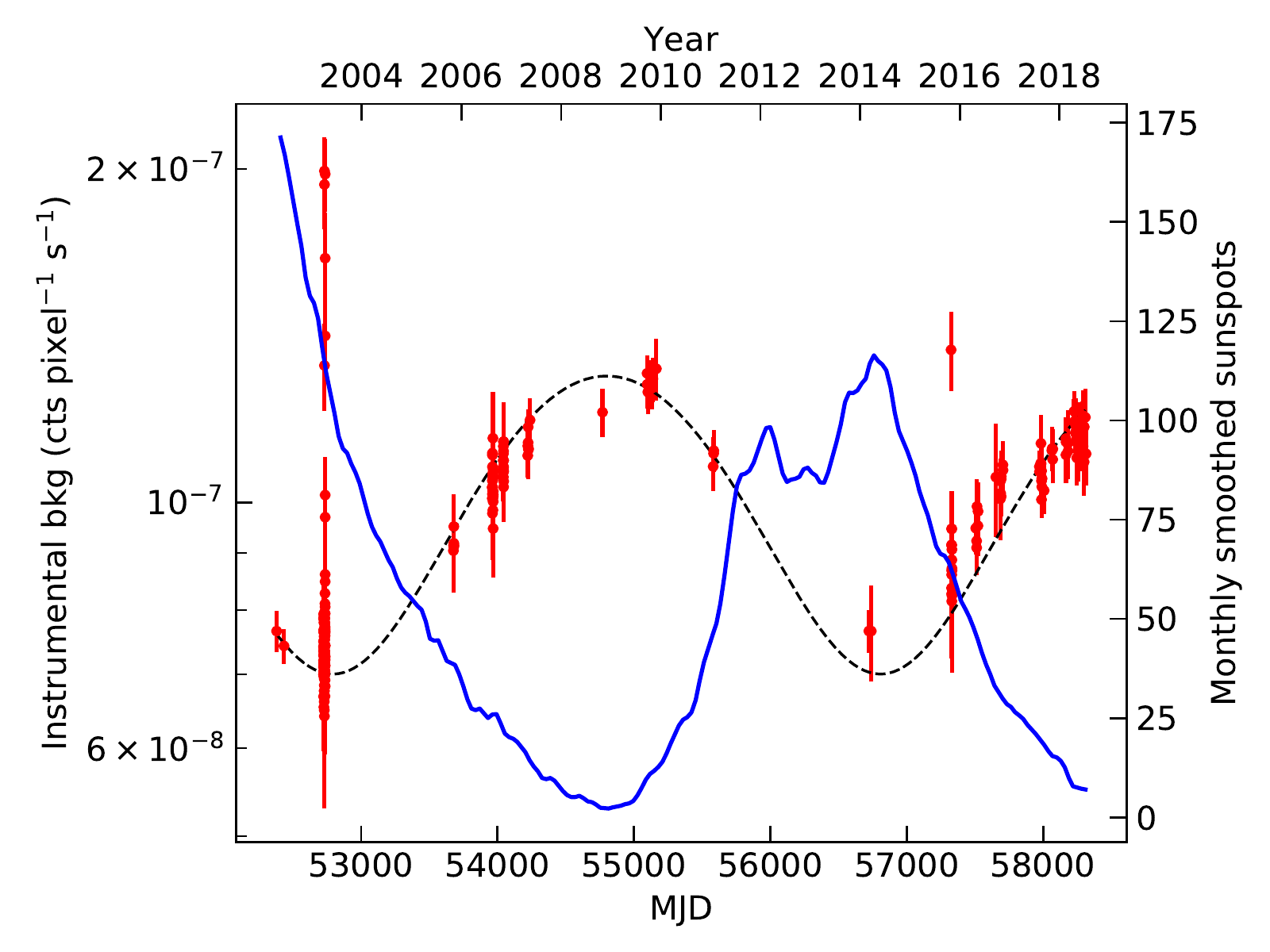}
    \caption{\chandra instrumental background surface brightness (in counts per pixel per second) as a function of modified Julian date (MJD, bottom axis) and, as a reference, year (top axis). The \xb observations during 2003 are seen on the left, where the small number of flared exposures as reported in the text and by \citet{kenter05} are visible as outliers. The outlier on the right is another flared short exposure ($\sim 6$ ks; ObsID 17423), which also was not discarded. The dashed black line marks a sinusoid with a period of 11 years, and it is not a fit to the data. It has been tuned to have a peak and a minimum at the approximate start and maximum of solar cycle 24, respectively. The solar activity is represented by the monthly smoothed sunspot number \citep[blue line,][]{sidc}, whose trend clearly shows an anticorrelation with the instrumental background surface brightness.}
    \label{fig:solarcycle}
\end{figure}

 We started by creating instrumental background maps following the procedure of \citet{hickoxmarkevitch06}. Briefly, since the \chandra effective area rapidly decreases at $E > 9$ keV, the $9-12$ keV band is vastly dominated by instrumental background. We then extracted the number of counts per pixel and per second in the $9-12$ keV band with \texttt{dmextract}. The \chandra instrumental background is well known to be anti-correlated with the solar cycle \citep[e.g.,][]{markevitch03}\footnote{The actual causes for the anticorrelation are complicated and still debated. The flux of particle background is likely modulated both by the solar wind speed and the solar magnetic field \citep[][and references therein]{rosschaplin19}: when the solar activity is higher, the cosmic rays modulation is driven by disturbances in the solar wind (and hence in the Heliospheric Current Sheet). It is likely that these disturbances result in a lower incidence of particle background on \chandra's FoV as well during solar maxima.}; thanks to the large time span of our observations ($\sim 15$ years), we were able to see this trend on more than one full solar cycle, as shown in Figure \ref{fig:solarcycle}. This confirmed also the vastly non-astrophysical nature of the events in the $9-12$ keV band: their number could thus be converted to instrumental background counts in other bands. \citet{hickoxmarkevitch06} estimated that the ratios between instrumental background in the F, S, and H bands and the one in $9-12$~keV are 1.52, 0.4, and 1.12, respectively. Using these ratios, the total number of counts in the F, S and H bands attributed to instrumental background could be obtained ($B_{\rm Instr}$), and compared to the total number of counts extracted from the data ($B_{\rm Data}$). 
To estimate the background directly from the data, for each observation we extracted photons from a $6\arcmin$-radius circular area centered on the aimpoint, appropriately masking out point sources detected in a given band with significance $> 3.5\sigma$ and clusters from the catalog of \citet{kenter05}. The counts extracted from the central $6\arcmin$ were then redistributed homogeneously over the whole FoV, rescaling by the ratio between the number of pixels of the ACIS-I FoV and those of the extraction area. Then, for each observation we compared the number of background counts extracted from the data with the counts estimated from the instrumental background alone: if the difference $B_{\rm Data} - B_{\rm Instr}$ was positively above $3\sigma$ (meaning that the data contained a significant number of excess counts of astrophysical nature), we subtracted the instrumental background from the total one, vignetted the resulting difference, and summed this component to the instrumental map to obtain the total background map ($B_{\rm Tot}$). Otherwise, the background from the data was considered consistent within $3\sigma$ with the instrumental background map previously created.
\par
\begin{figure}
    \centering
    \includegraphics[width=0.5\textwidth]{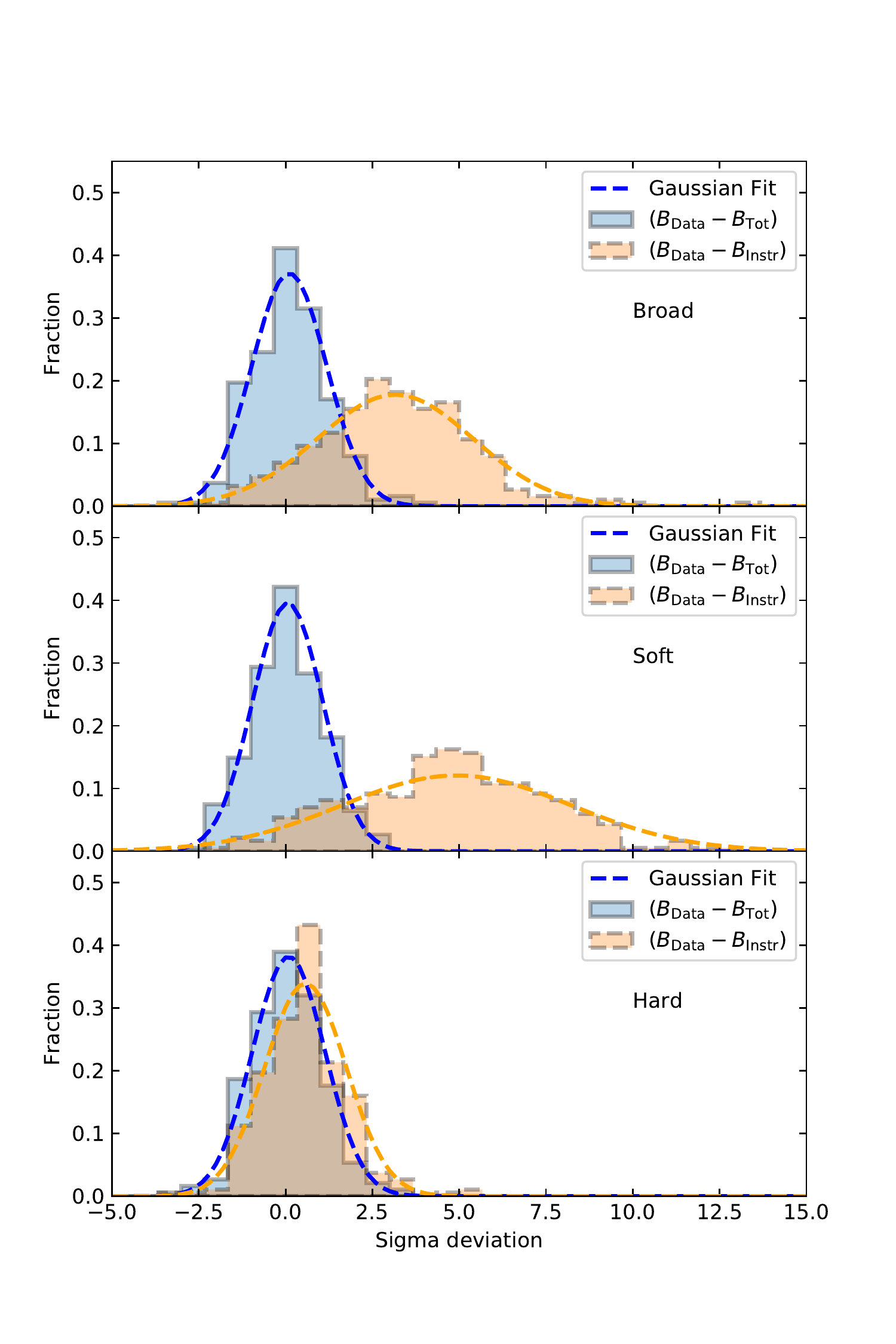}
    \caption{Distributions of the counts difference between $B_{\rm Data}$ and $B_{\rm Instr}$ (before applying any correction to our background maps, orange dashed histogram) and of the difference between $B_{\rm Data}$ and $B_{\rm Tot}$ (after adding the CXB and soft diffuse emission, light blue solid histogram). The dashed orange and blue lines are Gaussian fits to the two distributions. This figure shows that the final background maps we used in this work are consistent with the background extracted from the data.}
    \label{fig:bkgmap}
\end{figure}

 In Figure \ref{fig:bkgmap}, we show the distribution of the difference (in $\sigma$) $B_{\rm Data} - B_{\rm Instr}$ before applying any correction (orange dashed histogram), and of the difference $B_{\rm Data} - B_{\rm Tot}$ after taking into account the CXB and soft diffuse emission (light blue solid histogram), as well as their Gaussian fits (orange and blue dashed lines, respectively). As can be seen from Figure \ref{fig:bkgmap}, the blue histograms are centered on zero and are narrower than the orange ones. In the hard band the distributions are very similar, implying that the background is mostly instrumental and our extrapolation of the instrumental background from the $9-12$ keV band was able to explain the background in our data in the majority of the observations.  

\subsection{PSF Map Creation}

\chandra's PSF is known to be spatially and energy dependent. In particular, the farther away from the aimpoint and the higher the energy, the larger the size. The shape of the PSF also varies across the field of view (FOV) with the azimuthal angle, significantly deviating from a circular shape and becoming more and more elliptical. However, a usual way to approximate this behavior is by defining the radius that encircles $90\%$ of the energy, $r_{90}$. An approximate formula widely used in the literature \citep[e.g.,][]{hickoxmarkevitch06} based on the trend of the PSF size with off-axis angle $\theta$ shown in the \chandra Proposers' Observatory Guide\footnote{\href{http://cxc.harvard.edu/proposer/POG/pdf/MPOG.pdf}{http://cxc.harvard.edu/proposer/POG/pdf/MPOG.pdf}} is

 \begin{equation}
    r_{90} \approx
    \begin{cases}
      1'' + 10''(\theta/10')^2, & E = 1.5~\text{keV} \\
      1.8'' + 10''(\theta/10')^2, & E = 6.4~\text{keV}
    \end{cases}
\end{equation}

Using this approximate formula for $r_{90}$, we multiplied the computed value of $r_{90}$ in each pixel with the vignetting-corrected exposure map at the same pixel. We created maps of $r_{90} \times t_{\rm Exp}$ and $r_{90}^2 \times t_{\rm Exp}$, then merged all the observations together and divided the resulting mosaics for the total exposure map mosaic. This can be mathematically expressed (for $r_{90}$) as $\sum_{i = 1}^{N}{r_{90,i}t_{\text{Exp},i}}/\sum_{i = 1}^{N}{t_{\text{Exp},i}}$, where the sum is over the overlapping observations contributing to each pixel. This procedure effectively returned two mosaics of an exposure-weighted average of $r_{90}$ and $r_{90}^2$. While the former mosaic was used extensively in the following analysis to extract aperture photometry, the latter one was used to derive the sensitivity of the survey (\S\ref{sec:sensitivity}).

\subsection{Energy Conversion Factor Map Creation}

The large time span of our observations across 15 years of \chandra operation (between \chandra's Cycle 3 and 18), during which the spacecraft's effective area has significantly changed, implies that the same intrinsic flux, same spectral shape and the same exposure time will result in a different amount of counts detected in different positions on the field.
\par
\begin{figure}
    \centering
    \includegraphics[width=0.45\textwidth]{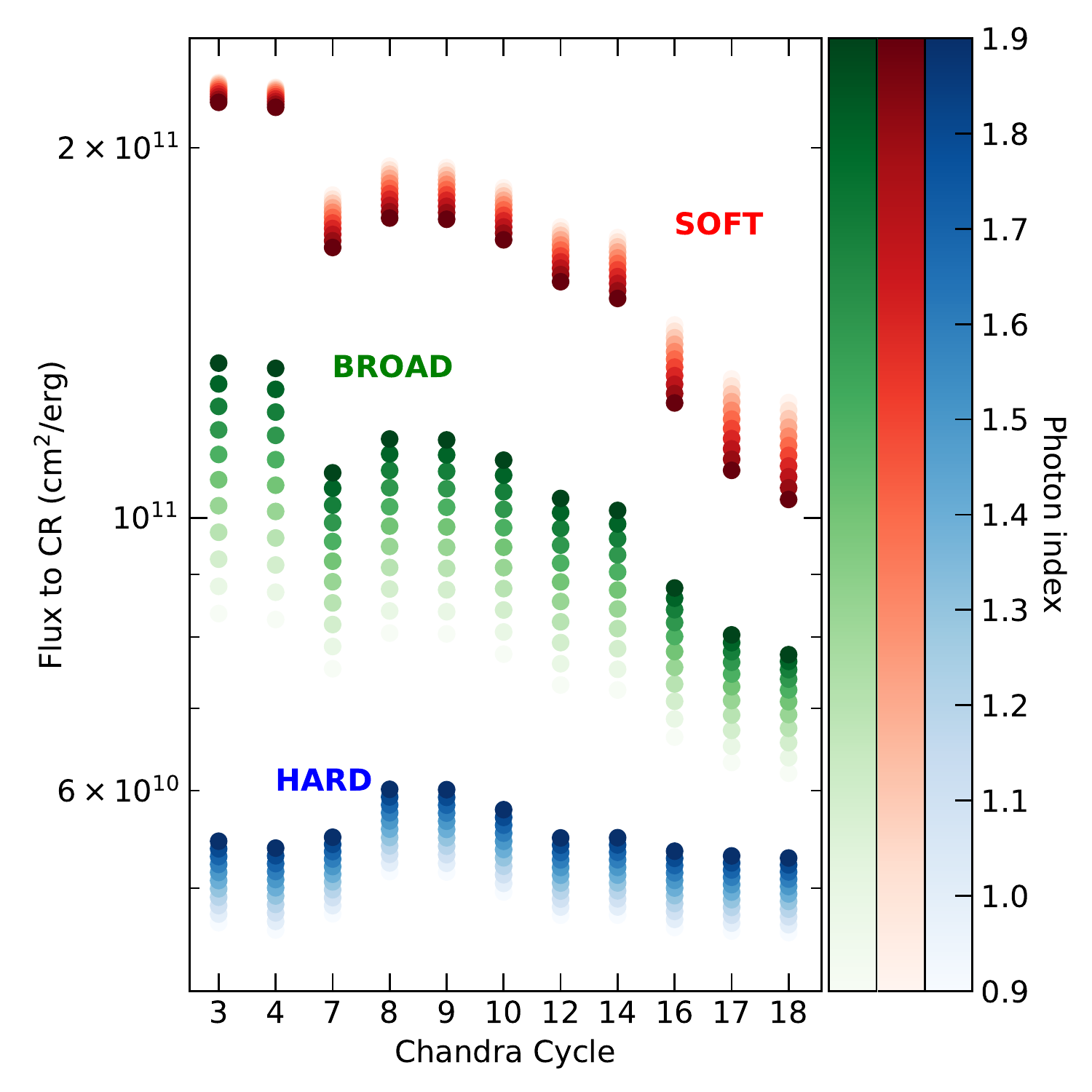}
    \caption{Energy conversion factors, from flux to count rate, as a function of \chandra Cycle, and color-coded by photon index. The green, red and blue colormaps refer to the F, S and H bands, respectively. Taking into account the temporal variation of the instrument is crucial for building accurate simulations.}
    \label{fig:ecfs}
\end{figure}

 We used PIMMS, as part of the \chandra Proposal Planning Toolkit\footnote{\href{http://cxc.harvard.edu/toolkit/pimms.jsp}{http://cxc.harvard.edu/toolkit/pimms.jsp}}, to obtain the energy conversion factors (ECFs) in the three energy bands adopted and for all the \chandra Cycles considered in this work (3, 4, 7, 8, 9, 10, 12, 14, 16, 17, 18), and for a range of photon indexes centered on $\Gamma=1.4$, which is the average photon index of the populations of AGNs mostly making up the unresolved CXB spectrum \citep{deluca04,hickoxmarkevitch06}. Figure \ref{fig:ecfs} graphically shows the significant evolution of the ECFs across the years, mainly in the soft band and seen regardless of the adopted spectral shape.
\par To create an ECF mosaic of the field, we first created the single maps. We considered, for each observation, the ECF given by its \chandra Cycle, assuming a power law with photon index $\Gamma = 1.4$, and Galactic absorption $N_{\text{H, Gal}} = 1.04 \times 10^{20}$ cm$^{-2}$  \citep{kalberla05}, and convolved the appropriate ECF with the vignetting-corrected exposure map. Then, analogously to what was done for the PSF maps, we merged the observations to create a mosaic, and divided the latter by the total exposure map mosaic.

\section{Simulations} \label{sec:sims}

In the original \xb survey, thanks to a remarkably uniform and low background, a simple cut in total counts could efficiently limit the number of spurious sources. When combining observations from different Cycles and with vastly different exposures, however, such a cut is not optimal, and the probability of being a spurious background fluctuation has to be carefully evaluated for each source. 
\par An effective strategy to approach this problem is to perform simulations of the observed data. To have full control of the numerous variables playing a role in a robust source detection, we decided to perform a full simulation of the entire \bootes field. 
\par The starting point is to assume a reasonable distribution for the number of sources per flux bin and per unit area (i.e. a differential number counts distribution, \dnds). We assumed the \citet{lehmer12} \dnds in the F, S and H bands separately, defined a range of fluxes (extending down to $\sim 5\times10^{-17}$ \fluxcgs, a factor of 20 lower than the expected flux limit), and made a Poissonian realization of the \dnds at each flux, creating the input list of sources. Each source was then assigned a random set of coordinates (R.A. and DEC.), accounting for the curvature of the sky\footnote{While the R.A. was assigned randomly, the randomly chosen DEC. was accepted only if $\cos(\text{DEC.}\times\pi/180) > P$, where $P$ is a random number uniformly drawn between zero and one. This (small) effect is introduced to more densely populate the low declination area of the field.}. Each source was also assigned a photon index $\Gamma$ between 0.9 and 1.9, drawn from a Gaussian probability distribution with $\mu=1.4$ and $\sigma=0.2$. When producing simulated sources, we did not account for the fact that the real sources might be clustered\footnote{The choice of the photon index range has a negligible impact on the result of the simulations, as we have tested that injecting sources with a fixed $\Gamma=1.4$ gives consistent results. This is not completely surprising, since the main goal of the simulations is to study the occurrence of background fluctuations as a function of probability threshold. For the same reason, we do not believe any missing clustering signal in the simulations to impact significantly our conclusions.}.
\par For each observation, we considered only those input sources falling on its FOV. Then, the coordinates of each source (R.A. and DEC.) were converted with the tool \texttt{dmcoords} into $\theta$ and $\phi$ (the distance in arcminutes of the given position from the aim point, and its azimuthal angle). These two quantities are crucial to take into proper account the effect of the variation of \chandra's PSF on the FOV. We used these coordinates to access the correct PSF image\footnote{The images of the PSF are stored in a file with a discrete range of elevations and azimuths, where elev $=\theta\sin{\phi}$ and azim $=\theta\cos{\phi}$.}, renormalize the image to have the expected number of counts given the input flux\footnote{Each source has its own randomly chosen photon index, which translates to a different ECF. In summary, we have a different ECF for any \chandra Cycle, any photon index, and any energy band employed.}, and place it on the instrumental background. The significant number of sources below the flux limit ensured that a similar effect to the CXB was naturally produced by undetectable sources. This procedure was performed for all the observations (adopting the same observational configuration for each), which were then merged together. The final simulation is then a Poissonian realization, pixel by pixel, of the whole mosaic.
\par Source detection was performed following a standard approach, extensively applied in the literature for many X-ray surveys \citep[e.g.,][]{nandra05,laird09,xue11,nandra15,civano16}. Sources were detected on the whole simulated mosaic with \texttt{wavdetect}, using the exposure map mosaic (with the parameter \texttt{expthresh} = 0.01) and the PSF map mosaic (the mosaic of exposure-averaged $r_{90}$) as additional inputs. We chose a permissive threshold \texttt{sigmathresh} $= 5\times10^{-5}$, analogous to what was done for \xb, and scales of ($\sqrt{2}$, 2, 4) pixels\footnote{We tested that adding an additional 8-pixel scale did not significantly change the results in terms of number of detected sources. The reliability thresholds determined through simulations were consistent with the ones we used in the main text, resulting in slightly fewer sources robustly detected.}. The source list returned by \texttt{wavdetect} was expected to contain many spurious sources due to the permissive threshold adopted, while, at the same time, being very complete.
\par We then performed aperture photometry at the position of each source returned by \texttt{wavdetect}, using the average $r_{90}$ at that position and extracting total and background counts, and the average vignetting-corrected exposure. Each source was assigned a probability of being spurious, i.e.\ the Poissonian probability that the observed total counts are entirely due to the expected background. Net (i.e.\ background-subtracted) counts ($N$) were then converted to count rates (CR $=1.1N/ t_{\rm Exp}$, where the factor $1.1$ corrects for the encircled energy fraction of the PSF) and finally to fluxes, through the ECF map mosaic -- remembering that the mosaics were created assuming a single photon index $\Gamma=1.4$. Uncertainties were computed following \citet{gehrels86}.
\par As a final step, the list of sources was cleaned for possible multiple detections of single bright sources, whose PSF wings can be independently detected. Feeding the PSF map mosaic to \texttt{wavdetect} helped to drastically reduce the number of such spurious detections (to less than $0.05\%$). When two sources were detected within the size of the PSF at that position, we retained the most significant one.
\par We have tested a wide range of simulations; altering the precise shape of the input \dnds, assigning the same and different photon indices to all the sources, changing the \texttt{wavdetect} input parameters, and running a set of ten realizations for each band and for each set. The results are consistent among each of the simulations. We also generated three independent lists of input sources, with a corresponding set of ten different Poissonian realizations. In the rest of this section, we discuss tests that use this total set of 30 simulations for each energy band.


\subsection{Setting a Probability Threshold}

As already mentioned, the list of sources returned by \texttt{wavdetect} was expected to be highly complete but, at the same time, include a significant number of spurious detections. Using a reasonable and justifiable probability threshold can drastically reduce the spurious fraction.
\par The total number of spurious sources detected in a simulation depends on Poisson statistics, on the magnitude of the background, and on the parameters of the detection algorithm (such as the \texttt{sigmathresh} and \texttt{scales} parameters of \texttt{wavdetect}). As long as the background used in the simulations is an accurate representation of the real background, and the detection process is the same as that adopted to detect sources in the real data, we can expect the number of spurious sources to be consistent within different realizations.
\par
\begin{figure}
    \centering
    \includegraphics[width=0.5\textwidth]{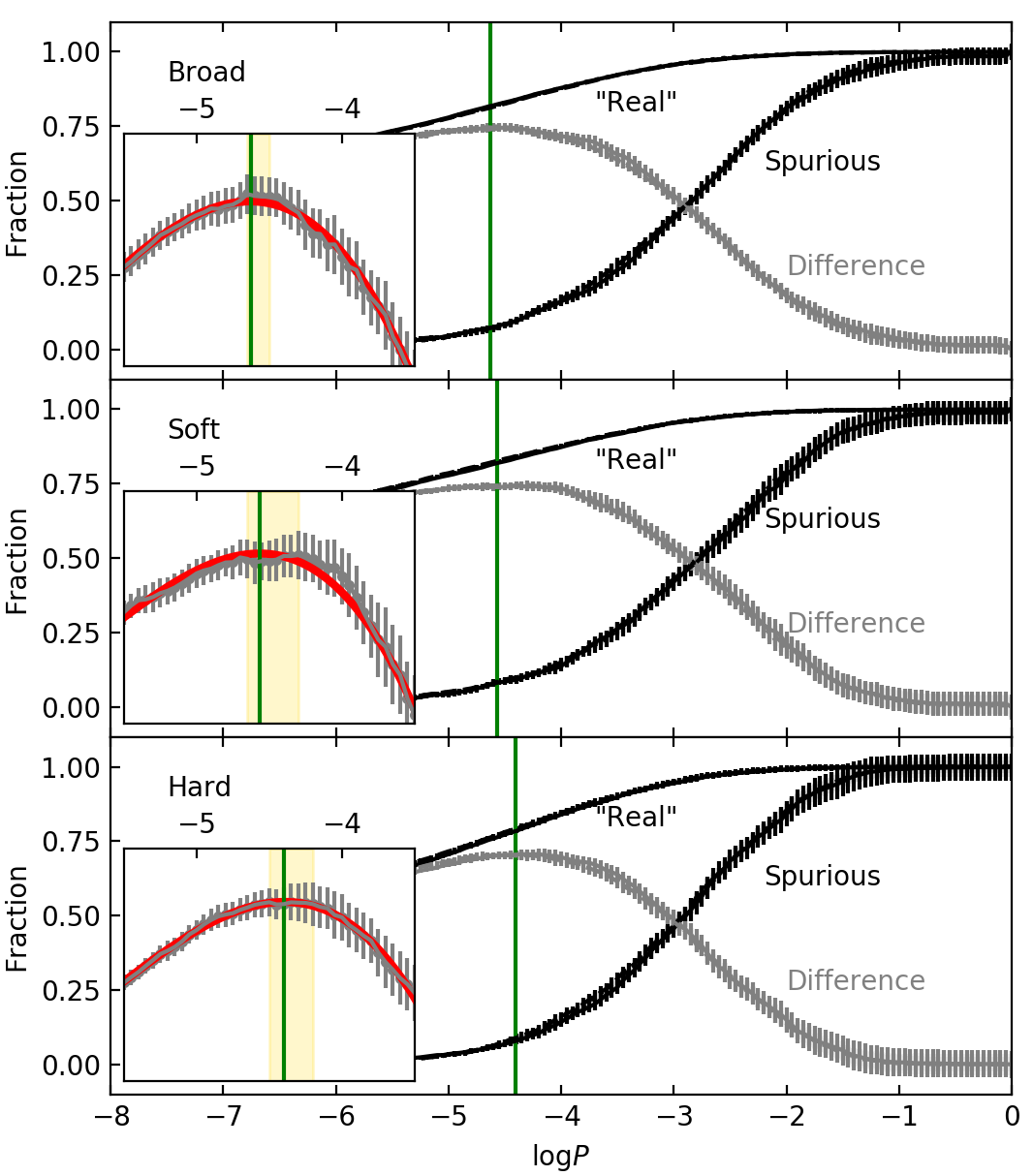}
    \caption{Distributions of the fraction of matched and not-matched sources for the three sets of simulations (black points and error bars) and their difference (gray points and error bars) as a function of logarithmic probability, $\log{P}$, for the three bands (from top to bottom, F, S and H). Each point is obtained by cutting the catalogs at the given $\log{P}$, and counting how many sources are/are not matched to their input counterparts. Cutting the catalog at higher $\log{P}$, more and more sources are detected but not matched to any input source. The green line labels the probability threshold that maximises the difference between matched and not-matched sources, obtained by fitting the peak range with a third-degree polynomial (red line in the insets, which show zoom-ins of the peak regions). The fitted positions of the peaks are consistent with the range defined by the single simulation sets (yellow region in the insets).}
    \label{fig:spurious}
\end{figure}

 Spurious sources are, by definition, sources returned by \texttt{wavdetect} that do not correspond to a ``real" input source. The fraction of spurious sources increases towards high probability (i.e., towards $\log{P}\sim0$), with a tail of fewer spurious sources being assigned a lower probability. We can use our simulated sources to compare the relative fraction of matched (i.e., ``real") and not-matched (i.e., spurious) sources at each probability. We expect these normalized distributions to have different shapes, so that we can determine a probability threshold that maximizes the difference between the two. This is shown in Figure \ref{fig:spurious}, where it can be seen that around $\log{P}\sim-4.5$ the difference between the two distributions is maximal. 
Due to statistical noise, the three sets of simulations (with different input sources) return slightly different peaks, as indicated by the yellow ranges in Figure \ref{fig:spurious}. Since setting an accurate probability threshold is crucial for this work, we computed the median of the three ``Difference" curves in Figure \ref{fig:spurious}
and fitted it with a third-degree polynomial function over the range $-6 < \log{P} < -3$. The fits are shown in the insets in Figure \ref{fig:spurious}.
The final thresholds we determined to minimize the selection of spurious sources are
$\log{P} = (-4.63, -4.57, -4.40)$ for the F, S and H bands, respectively.

\subsection{Completeness and Sensitivity}\label{sec:sensitivity}

\begin{figure}
    \centering
    \includegraphics[width=0.49\textwidth]{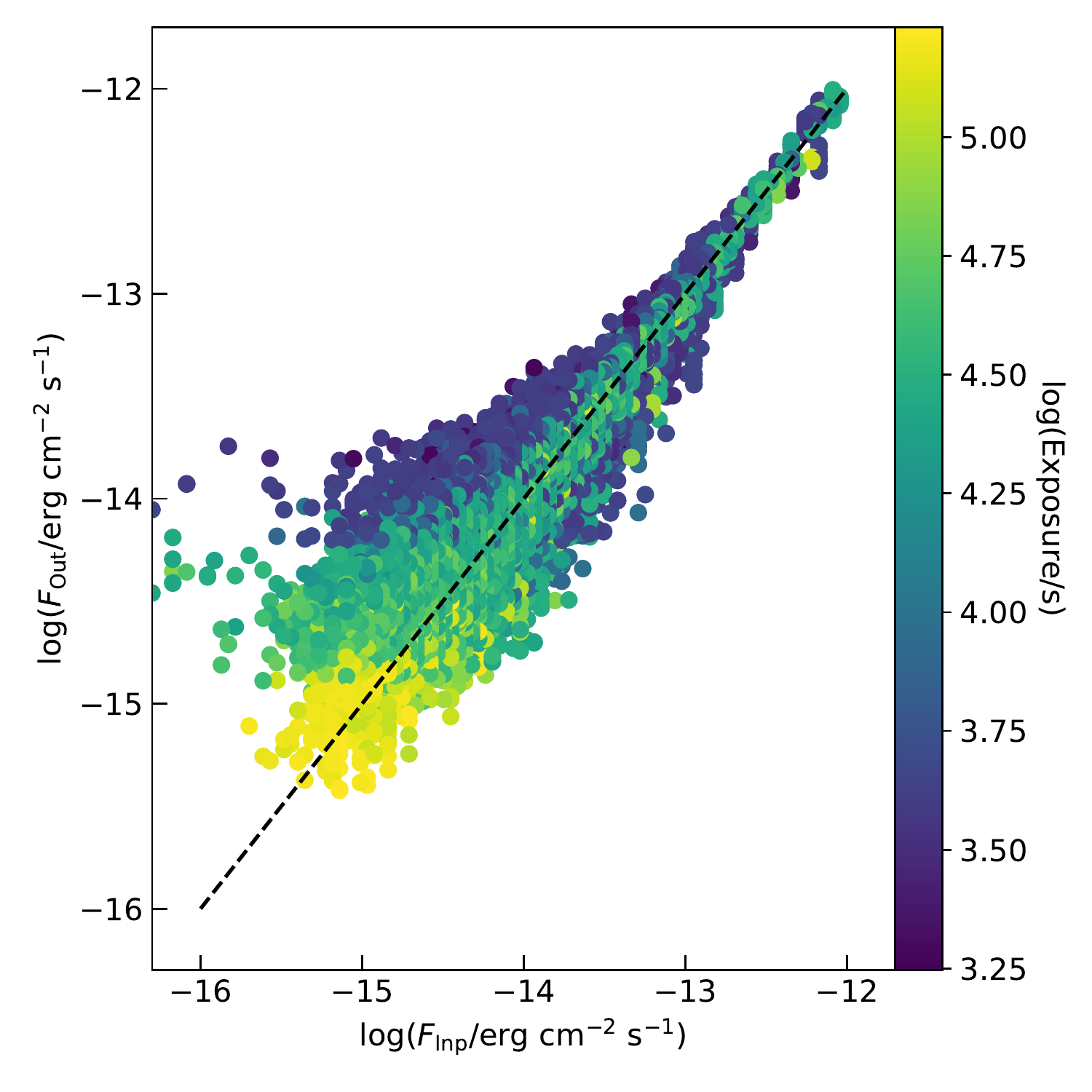}
    \includegraphics[width=0.49\textwidth]{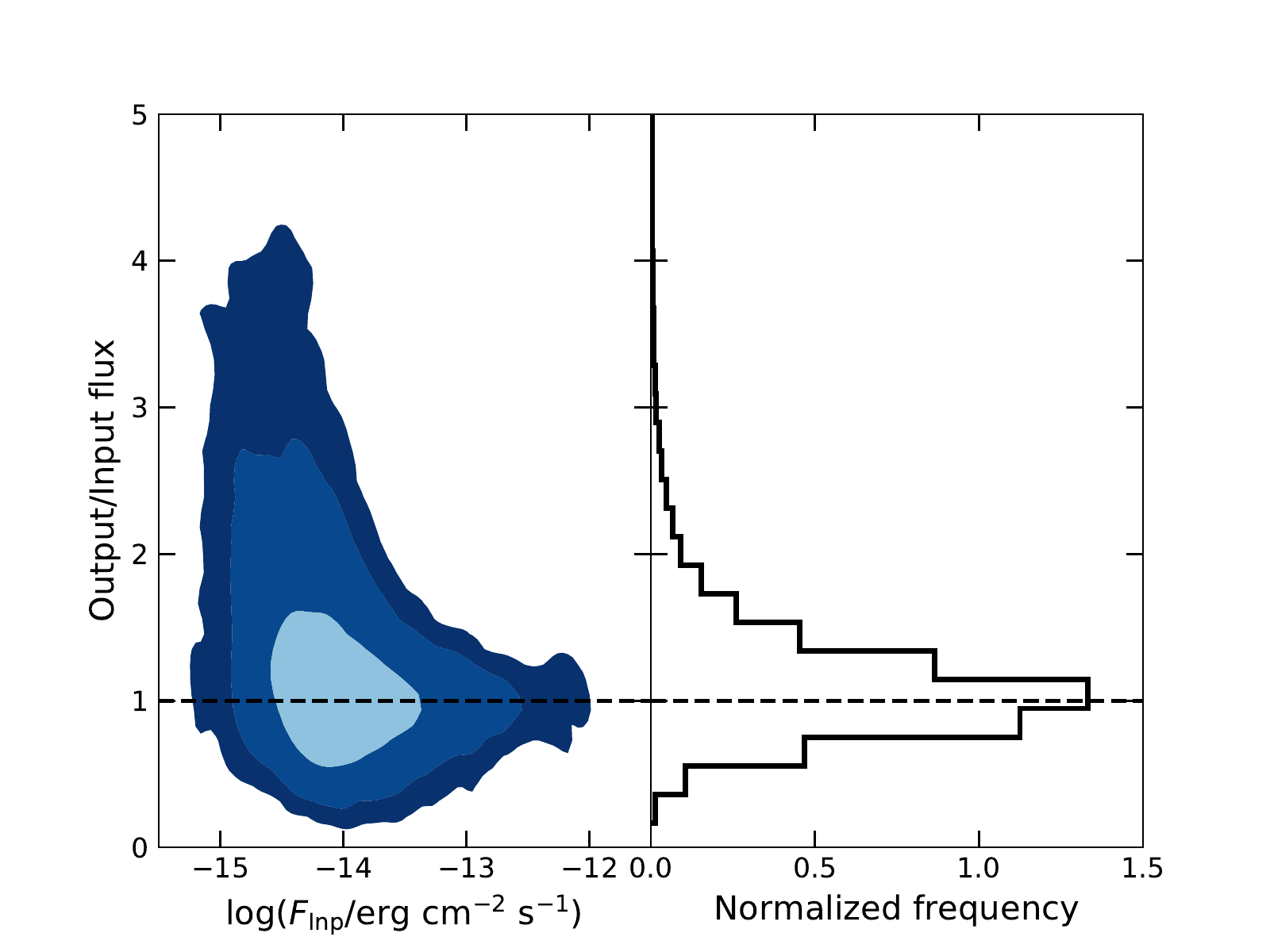}
    \caption{\textbf{Top.} Comparison of the output flux (measured through aperture photometry) as a function of the relative input flux, for sources simulated, detected and matched to their input counterparts, for a full set of ten realizations. The dashed black line shows the 1:1 relation. The Eddington bias effect is clearly visible at fluxes $F \lesssim 10^{-14}$ \fluxcgs, as an increased systematic positive spread in the output fluxes toward faint input fluxes. The points are color-coded by the logarithm of their exposure time, with deeper exposures being labeled by brighter colors. \textbf{Bottom.} On the left, the contours of the ratio between the output fluxes over input fluxes, as a function of input flux. The contours label the $68\%$, $95\%$, and $99\%$ of the sources, respectively, in shades of darker blue. On the right, the same distribution showing a sharp peak around 1 (black dashed horizontal line) and the Eddington bias tail.}
    \label{fig:fin-fout}
\end{figure}

Using the simulations, we could test how well we measure fluxes, comparing output and input fluxes, and compute the completeness of our sample. In particular, the comparison of the output and input fluxes shows the well known Eddington bias \citep{eddington13}, for which faint sources, close to the flux limit of the survey tend to be brighter than their actual flux, due to positive fluctuations being more likely to be detected than negative ones. This is exemplified in the top panel of Figure \ref{fig:fin-fout}, where the points show (detected) single sources matched to their input counterparts. The Eddington bias effect is also clearly visible in the bottom panel of Figure \ref{fig:fin-fout}.
\par
\begin{figure*}
    \centering
    \includegraphics[width=\textwidth]{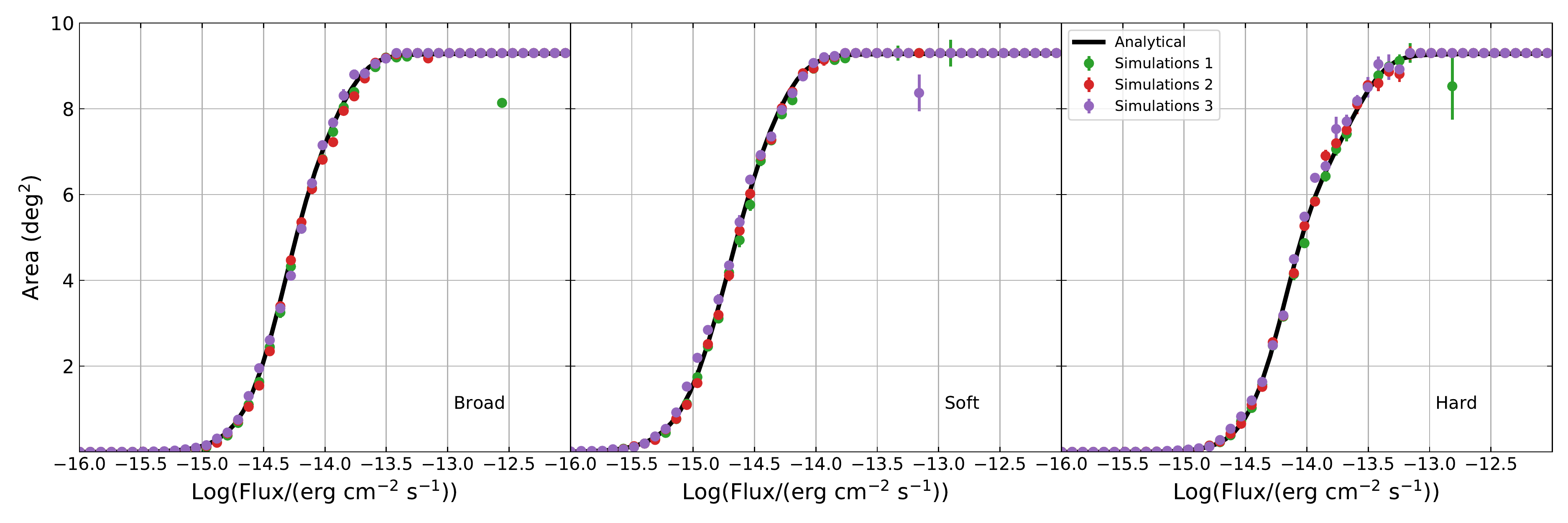}
    \caption{Sensitivity of the CDWFS survey for the bands employed (from left to right: broad, soft, hard). The results from our three independent sets of simulations (each set has a different list of input sources) are shown with colored points (median of ten simulations), with standard deviation as uncertainty. The solid black line labels the sensitivity computed following \citet{georgakakis08}, whose analytical computation includes the Eddington bias and expected spurious sources. The dips at bright fluxes are mostly due to bright sources being undetected due to edge effects; see the text for more details.}
    \label{fig:sens}
\end{figure*}

 The ratio of the number of matched sources to the number of input ones, as a function of (input) flux, decreases towards fainter fluxes, reflecting the incompleteness of the survey. This ratio, rescaled by the total area of the survey, is in excellent agreement with the expected sensitivity curve obtained using an analytical calculation that includes Eddington bias \citep{georgakakis08}, as shown in Figure \ref{fig:sens}.
To compute the expected sensitivity curve taking into account Eddington bias, we first multiplied each pixel of the background map mosaic with the exposure-averaged area of the PSF at that location, given by $\pi r_{90}^2$, so that $B^* = B\times\pi r_{90}^2$. Then we computed the minimum number of counts $cts$ that, given the background value $B^*$ in that pixel, would result in a significant detection (i.e., would result in a probability that $cts$ are due to a background fluctuation, lower than the imposed threshold). Then, we defined an array of fluxes $s$, and we converted it into an array of expected number of counts, given by $T = s\mathcal{C}f_{\rm PSF}t_{\rm Exp} + B^*$, where $\mathcal{C}$ is the energy conversion factor from flux to count rate, $f_{\rm PSF}$ is the encircled energy fraction of the PSF (0.9 in our case), and $t_{\rm Exp}$ is the exposure time. For each pixel, we ended up with an array of probabilities $P(cts,T)$ that the minimum number of counts $cts$ are observed, given the expected number of counts $T$. Finally, summing up all the pixels at any given flux resulted in the expected sensitivity curve.
The dips at high fluxes in Figure \ref{fig:sens} are due to a handful of bright sources being undetected. Such sources mainly fall on the very edge of the mosaic and have a sufficiently large PSF that their effective surface brightness is lowered. In one case, a bright source was missed entirely because it lied very close to another (brighter) source and was detected and rejected by our code that removes duplicate detections in PSF wings. Representative values of flux limits at different levels of completeness are tabulated in Table \ref{tab:completeness}.

\begin{deluxetable*}{lccccccccccc}
\tablecaption{Flux limits at different completeness values. \label{tab:completeness}}
\tablehead{\colhead{Band} & \multicolumn{11}{c}{Completeness} \\
\cline{2-12}\colhead{} & \colhead{99\%} &\colhead{95\%} & \colhead{90\%} & \colhead{80\%} & \colhead{70\%} & \colhead{60\%} & \colhead{50\%} & \colhead{40\%} & \colhead{30\%} & \colhead{20\%} & \colhead{10\%}} 

\startdata
F & 33.3 & 20.6 & 15.7 & 11.0 & 8.26 & 6.62 & 5.45 & 4.51 & 3.71 & 2.96 & 2.14 \\
S & 13.1 & 7.86 & 5.95 & 4.20 & 3.24 & 2.59 & 2.12 & 1.74 & 1.41 & 1.10 & 0.77 \\
H & 66.1 & 40.5 & 30.7 & 20.4 & 14.1 & 10.6 & 8.39 & 6.91 & 5.70 & 4.58 & 3.35
\enddata

\tablecomments{Fluxes are in units of $10^{-15}$ \fluxcgs. To obtain the area covered at any tabulated flux limit, the completeness must be rescaled by the total area of the field, $9.3$ deg$^2$.}
\end{deluxetable*}

\section{Source detection} \label{sec:sdet}

The same detection procedure used for the simulations was applied on the real data mosaics, separately for the three bands. After running \texttt{wavdetect} on each data mosaic separately, we refined the positions of each candidate source following \citet{murray05}. Since the mosaics have a pixel scale of $1\farcs968 \times 1\farcs968$\footnote{There could be cases in which multiple nearby sources, with a PSF smaller than the $4\times4$ pixel scale, were detected as one single source. Hence, we visually inspected all the 562 sources in the final catalog with $r_{90} < 1\farcs96$, finding only one case of two sources blended as one due to the spatial binning. We corrected the entry of this source, adding two distinct entries at the bottom of the final catalog.}, we produced a full resolution cutout of 100 native \chandra pixels around the position of the putative source, and iteratively centroided the events within $r_{90}$\footnote{In the case of sources for which $r_{90}$ is smaller than 3\arcsec, we fixed it to $3\arcsec$ to avoid the rare occurrence of small regions with zero events to centroid.}.  Once the coordinates were refined for the whole list of candidate sources returned by \texttt{wavdetect}, we performed aperture photometry on the full-resolution mosaic, and cleaned each list for duplicates.
\par Since a large fraction of these sources are expected to be detected in multiple bands, a cross-match between the F, S and H catalogs was done using $r_{90}$ as a matching radius. The merging of the catalogs returned \beforexb unique X-ray sources detected above the reliability threshold in at least one band. The average number of spurious sources detected in the simulations (58, 48.5, 52) translates into an expected spurious fraction of $\sim (0.9\%, 0.9\%, 1.6\%)$ for the F, S and H bands, respectively.

\subsection{Positional Errors}

\begin{figure}
    \centering
    \includegraphics[width=0.5\textwidth]{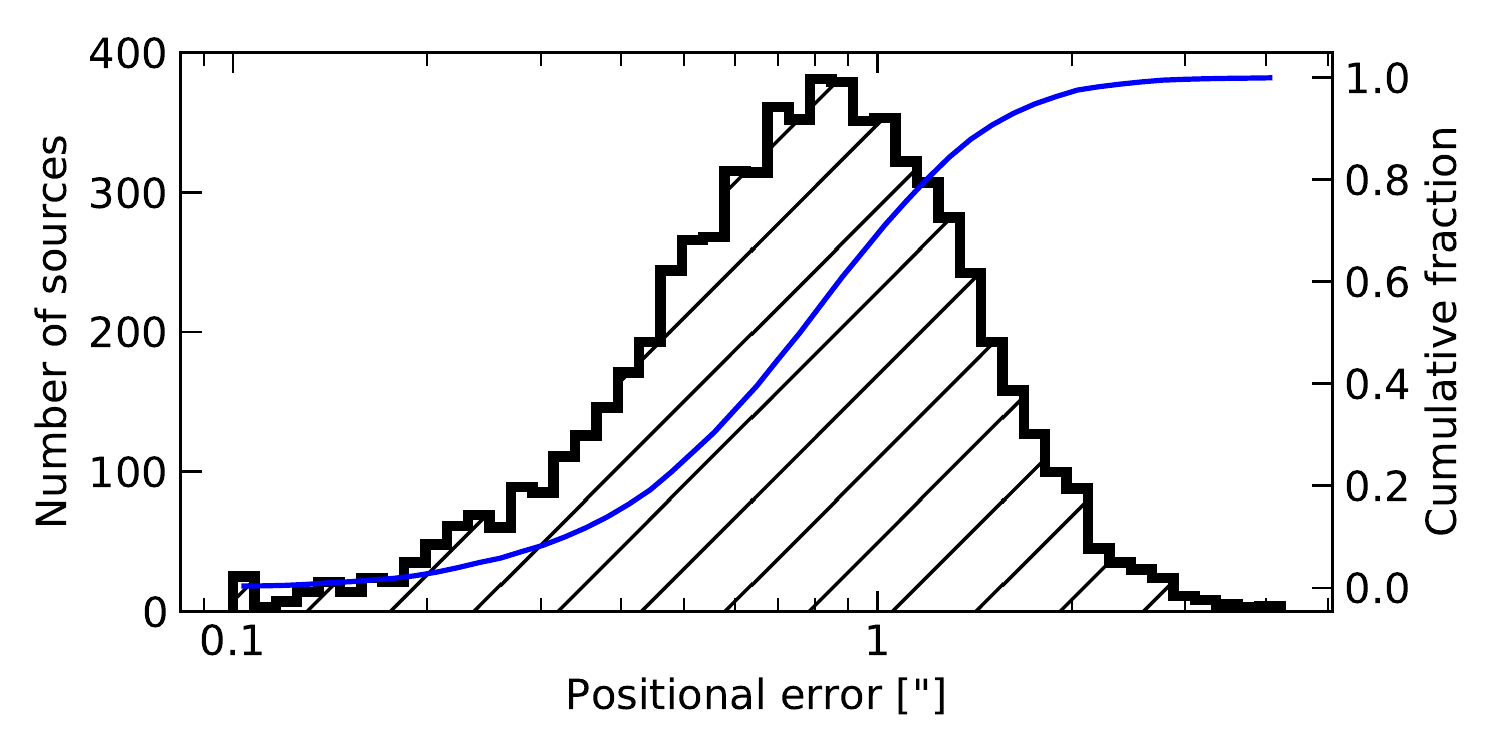}
    \caption{Distribution of the X-ray positional errors for our sample, computed following \citet{puccetti09}. The blue line shows the cumulative histogram, and implies that $90\%$ of the sources have an uncertainty on their position lower than $1.5''$.}
    \label{fig:poserr}
\end{figure}

When a source was detected in more than one band, we used the coordinates of the band with the most significant detection as the final X-ray position, and we computed the positional error as $r_{50}/\sqrt{N}$ \citep{puccetti09}, where $r_{50}$ was estimated from $r_{90}$ as $r_{50} = 5/9\times r_{90}$\footnote{We verified that this formula correctly approximates the increase of PSF size with off axis angle, as shown in the \chandra Proposers' Observatory Guide.}, and $N$ are the net counts (within $r_{50}$) in the band with the most significant detection. When $r_{50}$ was smaller than a $4\times4$ pixel area of our mosaic, we used $r_{90}$ instead (this was necessary for $\sim 2\%$ of the sources). The distribution of the positional errors derived in this way is shown in Figure \ref{fig:poserr}, and has a median value of $\sim 0\farcs8$. Positional errors formally smaller that $0\farcs1$ (occurring for 19 very bright sources) have been conservatively set to $0\farcs1$, following \citet{civano16} and \citet{puccetti09}. The cumulative distribution shows that $90\%$ of the sources have a positional error $<1\farcs5$ and $99\%$ less than $2\farcs5$, while $71\%$ of them have a positional error smaller than $1\arcsec$. These somewhat non-optimal values \citep[compared, for example, with the values reported by][for the \chandra COSMOS Legacy survey, in which $85\%$ of the sources have a positional error $<1\arcsec$]{civano16} could be attributed to the large fraction of observations overlapping with a significantly different roll angle, which ultimately results in a slightly larger PSF size on some parts of the field, and to the lower median net counts of the CDWFS sources (15, 11, and 13 compared to 30, 20 and 22 in \chandra-COSMOS Legacy in the F, S and H bands, respectively). Our positional errors are consistent with the errors derived using 68\% confidence level formulas as reported by \citet{kim07}. Hence, we consider our positional errors as $1\sigma$ uncertainties on the position of detected sources.

\subsection{Counts and Fluxes}

Once a unique set of coordinates was derived for each source, we re-extracted aperture photometry for all the bands at the same position, analogous to what was done for the simulations. We briefly recall the procedure here: using the exposure-weighted average $r_{90}$ to extract counts, background counts and exposure, we computed the net counts and converted them to a count rate. The flux was computed correcting the count rate for the encircled energy fraction of the PSF within $r_{90}$, and converted to flux using the exposure-weighted average ECF (assuming $\Gamma = 1.4$). Using a single set of coordinates for each source, the most significant detection had exactly the same aperture photometry as before, while the other two bands were adjusted and realigned with the new coordinates.
\par This procedure assigned, for the bands in which each source was not significantly detected, a new probability of being spurious. If this new probability was higher than our imposed detection threshold, we considered the source to be detected in that band (although originally missed by \texttt{wavdetect}), and computed its count rate and flux as outlined above. If, instead the source had a probability of being spurious higher than our adopted threshold, we computed $3\sigma$ upper limits on the net counts, count rate and flux, following \citet{gehrels86}.
\par
\begin{figure}
    \centering
    \includegraphics[width=0.5\textwidth]{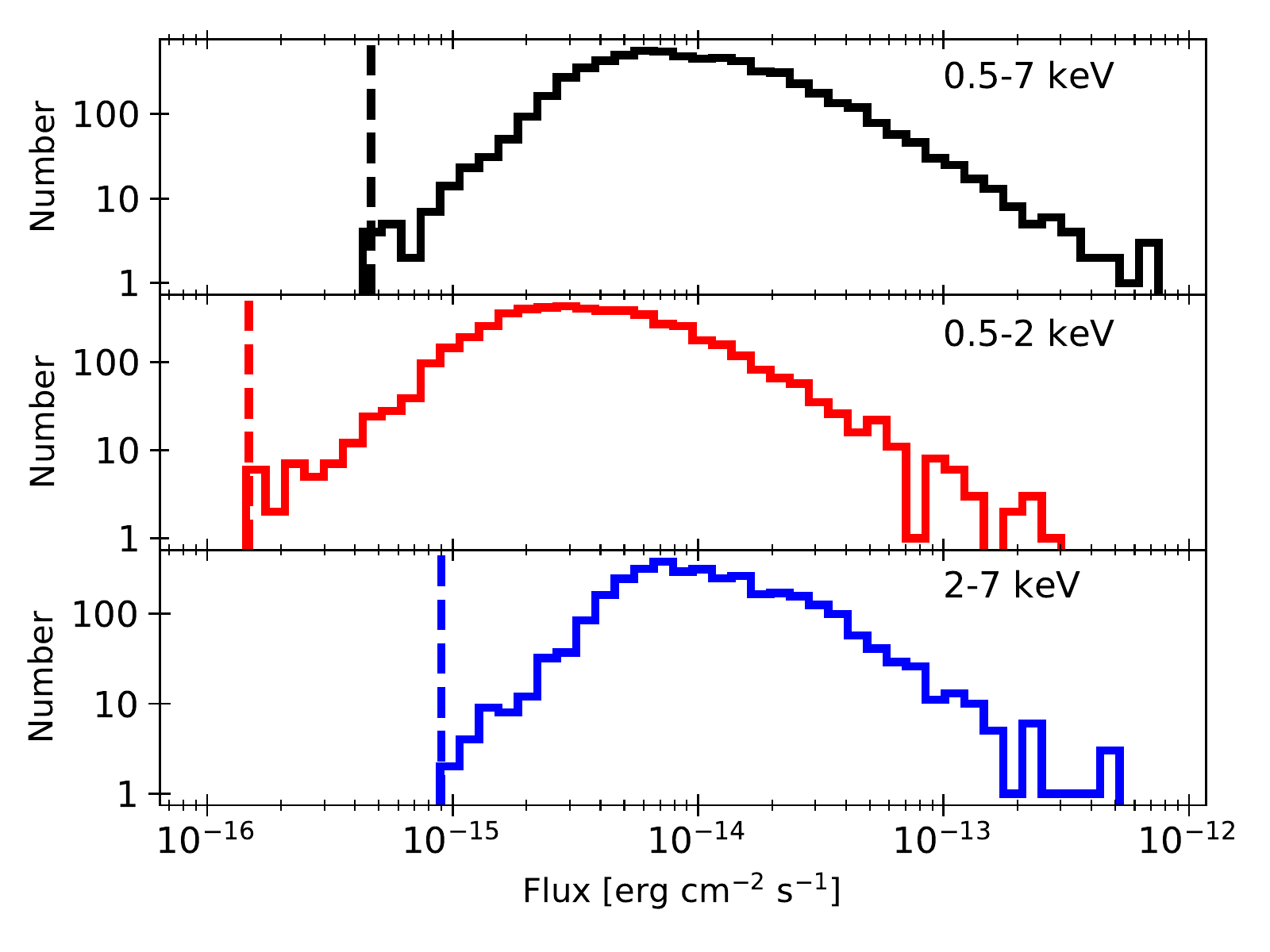}
    \caption{Distribution of the X-ray fluxes in the three bands as labeled (from top to bottom: broad, soft, and hard bands, respectively), computed with aperture photometry. In each panel, the vertical dashed line marks the faintest flux with a significant detection. We did not include upper limits for sources undetected in each band.}
    \label{fig:fluxes}
\end{figure}

 The distributions of fluxes measured for sources significantly detected in each of the three bands are shown in Figure \ref{fig:fluxes}; for display purposes, the brightest source in the field, i.e. the variable star KT Bo\"otes (F-band flux of $F_{\rm X} \sim 3 \times 10^{-12}$ \fluxcgs), is not included.

\subsection{Comparison with \xb}

As a final step, we cross-matched our catalog with that of \citet{kenter05}, which contains 3293 X-ray point sources detected with more than 4 counts\footnote{Due to the astrometric shift discussed in \S \ref{sec:data_red} and in the Appendix \ref{asec:astrometry}, we shifted the position of each \xb source before matching it to our catalog.}. We found that \xbmissed out of 3293 \xb sources ($\sim 14\%$) do not have an entry (within $1.1\times r_{90}$) in the CDWFS catalog. Such a high number of missing sources cannot be simply explained by the spurious fraction of the \xb survey, expected to be around 1\% \citep{kenter05}. The large majority (95\%) of the missing \xb sources have $\leq 6$ counts in the \citet{kenter05} catalog. Of the 21 missing sources with 7 or more counts, half of them are spread on such a large PSF area that the background is non-zero, and their significance drops. Other few sources are genuinely drown into the background with deeper data, for others there is a counts mismatch, i.e. unrealistically large PSF radii are required to match the number of counts reported in the \xb catalog. These latter cases imply that the updated data reduction and/or different light curve filtering may have introduced some subtle differences among the CDWFS and \xb data.

\begin{deluxetable}{lcccc}
\tablecaption{Breakdown of the missing \xb sources. \label{tab:xb_missing}}
\tablehead{\colhead{\multirow{2}{*}{}} & &  \colhead{\texttt{wavdetect}} & \colhead{Reliability} & \colhead{Total}} 

\startdata
\multicolumn{1}{l}{Not significant} & & 243 & 151 & \xbbelowthresh \\\cmidrule{1-1}
\multirow{2}{*}{Exposure} &  $<10$ks &  75 & 104 & 179  \\ 
                          &  $>10$ks &  168 & 47 & 215  \\
\midrule 
\multicolumn{1}{l}{Significant} & & 40 & 13\tablenotemark{a} & \xbtba \\\cmidrule{1-1}
\multirow{2}{*}{Exposure} & $<10$ks &  14 & 4 & 18  \\ 
     				      & $>10$ks &  26 & 9 & 35  \\ 
\midrule
 Total missing  & & 283 & 164 & \xbmissed \\ \bottomrule
\enddata

\tablecomments{The column \texttt{wavdetect} refers to \xb sources missing from the whole list of candidate sources in output from \texttt{wavdetect}, while the column Reliability refers to the additional missing sources excluded by our imposed probability cuts. Of the \xbmissed missing \xb sources, the \xbtba significant ones were included in the CDWFS catalog, while the \xbbelowthresh not significant were discarded. However, an additional table is provided listing all the \xbbelowthresh excluded \xb sources (see Appendix \ref{asec:xbootes}).}
\tablenotetext{a}{The very few sources missing due to reliability cuts, but then re-added for being significant, are sources for which a tiny shift between the candidate position returned by \texttt{wavdetect} and the actual \xb position is enough to move the source above/below the threshold.}

\end{deluxetable}

\begin{figure*}
    \centering
    \includegraphics[width=0.47\textwidth]{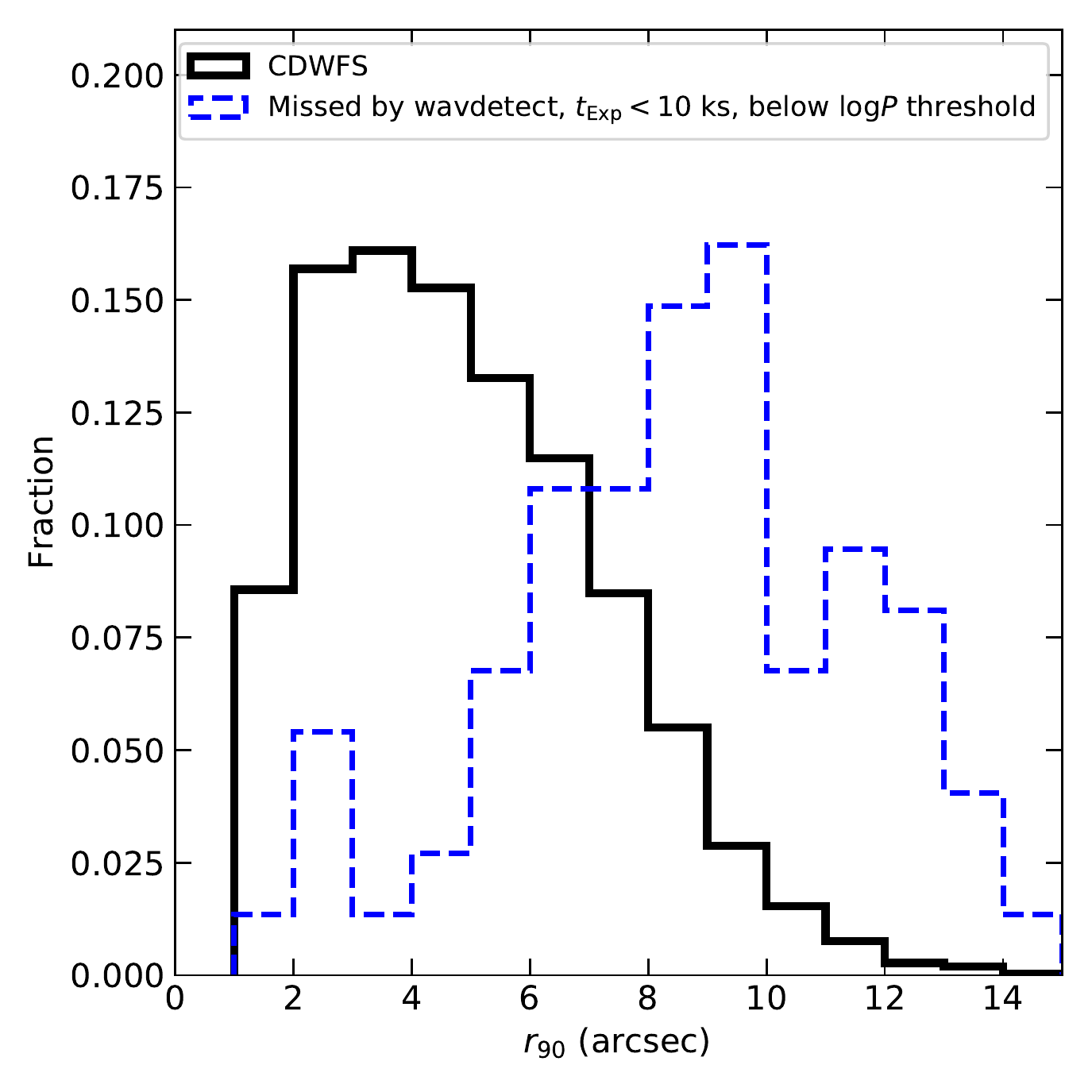}
    \includegraphics[width=0.47\textwidth]{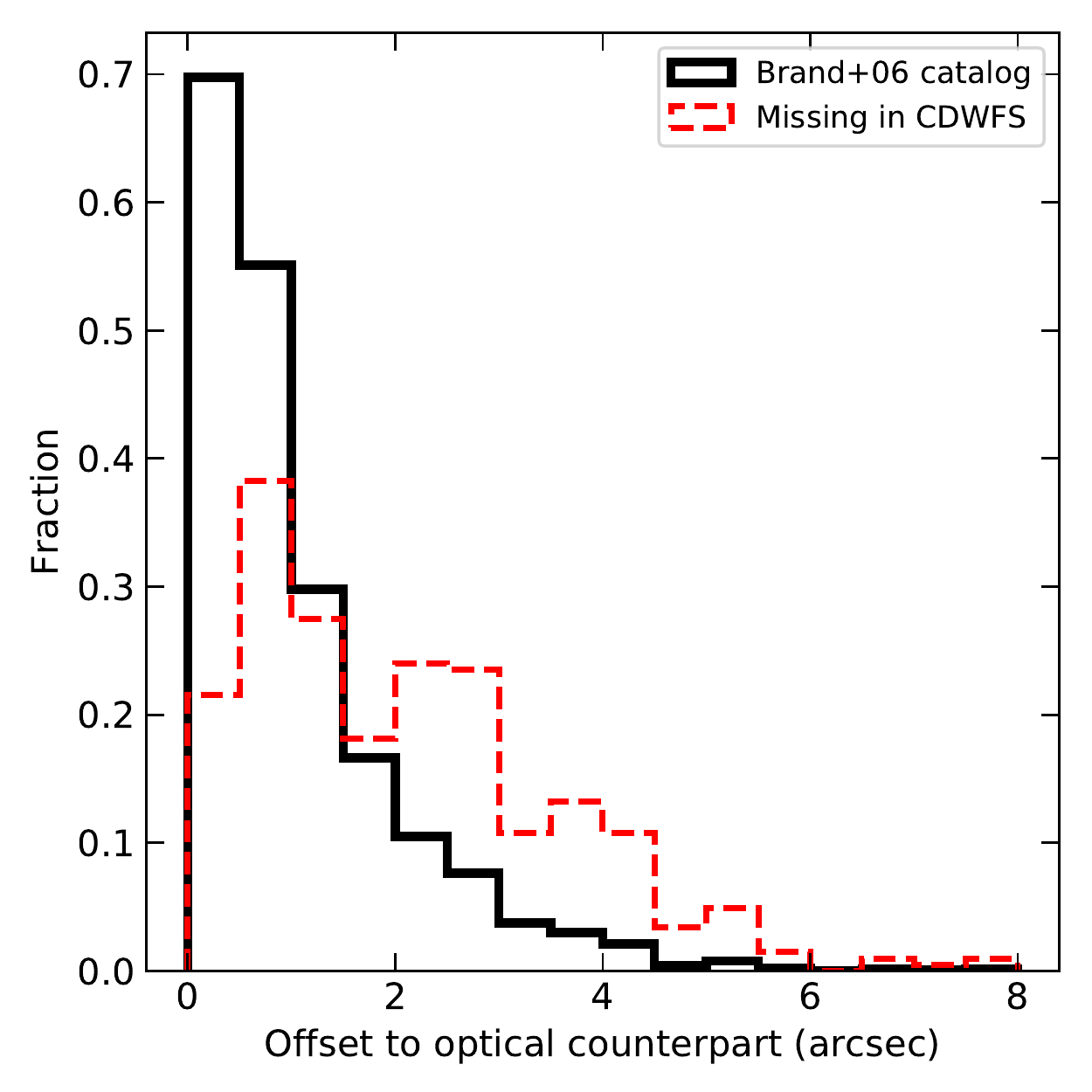}
    \caption{\textbf{Left.} Normalized distribution of the average $r_{90}$ for the full CDWFS sample (solid black histogram) and the subset of \xb sources missing from the output list of candidate sources having exposure $<10$ ks (dashed blue histogram). The two distributions are clearly different and suggest that \texttt{wavdetect} does not detect such sources due to the additional PSF size information, differently to what was done in \citet{murray05}. \textbf{Right.} Normalized distribution of the offset between \xb sources and their optical counterparts as measured by \citet{brand06}. The missing sources (red dashed histogram) are on average more offset from their best (i.e., higher rank) optical counterparts than the full catalog, labeled by the solid black histogram. This may indicate that the positional error of the missing sources is larger than average, due to their low counts and large off-axis angles.}
    \label{afig:xb_offset}
\end{figure*}


\par Table \ref{tab:xb_missing} shows the breakdown of the missing sources, split into sources absent from the candidate list of sources in output from \texttt{wavdetect}, and those additionally missed due to our reliability cuts (i.e., detected as potential sources by \texttt{wavdetect} and later rejected). It can be seen from Table \ref{tab:xb_missing} that the majority of the missing sources ($283/447 \sim 63\%$) are already missing from \texttt{wavdetect} output of candidates. While $2/3$ of the total area of the field is covered by additional data compared to the original \xb survey, $1/3$ of the area ($\sim 3$ deg$^2$) comprises the very same data. A rough distinction between sources lying on the same \xb data and those falling where new observations have been done, is obtained in Table \ref{tab:xb_missing} by splitting the missing sources by exposure: sources with $t_{\rm Exp} < 10$ ks lie mostly over the original \xb edge of the field.
\par While Table \ref{tab:xb_missing} reveals that the sources missed by \texttt{wavdetect} are homogeneously scattered across the field regardless of exposure time, the reasons for missing sources in areas of different exposure are likely different. On one hand, when the exposure time is increased by co-adding observations, the Eddington bias effect becomes less significant, source variability can play a role, and genuinely spurious sources are less likely to be detected again. Among these factors, the first one is likely the most effective in this case: our simulations show that $\sim 16\%$ of simulated sources with $t_{\rm Exp} < 8$ ks are detected with an output flux more than twice the input one, due to Eddington bias. The effect drastically reduces to $\sim 3\%$ for sources with $t_{\rm Exp} > 40$ ks. A possible additional role could be played by the degrading sensitivity of \chandra with time, resulting into diluting the genuine signal of faint sources with the increasing exposure (and hence background).
\par On the other hand, \xb sources missing from \texttt{wavdetect} output on the very same \xb data require a more careful treatment and a different explanation. As discussed in \S \ref{sec:sims}, \texttt{wavdetect} was run on the whole field mosaic, with additional inputs such as the exposure map and PSF mosaics\footnote{The scales over which \texttt{wavdetect} was run are also slightly different compared to \xb; we did not include the largest 8 pixels scale (corresponding to $8\times1\farcs968 = 15\farcs74$) mentioned by \citet{kenter05}, but we verified that the difference cannot be explained by adding this scale.}. \citet{murray05} and \citet{kenter05} do not mention if any additional information was fed to \texttt{wavdetect}, but it is likely they did not use exposure maps nor PSF maps when detecting sources over each of the 126 \xb observations. In the left panel of Figure \ref{afig:xb_offset} can be seen that our inclusion of the PSF information is very likely playing a role. The median $r_{90}$ of the sources lying on the low-exposure part of the field and missed by \texttt{wavdetect} is almost twice the median $r_{90}$ of the CDWFS catalog\footnote{The effect is seen also with the total sample of sources missed by \texttt{wavdetect}, although with a lower significance.}. Smoothing the low counts of such sources on a large PSF area likely dilutes their contrast with the background and lowers their significance. This emphasizes a caveat regarding using a simple circular area of radius $r_{90}$ to detect sources and to extract aperture photometry. The sampling of the PSF is indeed not homogeneous, and counts (especially in low-exposure, low-counts regime) will generally not distribute uniformly across the circular area. A more rigorous approach would require taking into account the actual shape and substructure of the PSF when detecting sources and extracting aperture photometry, which is outside the scope of this work.
\par The fact that sources missed by \texttt{wavdetect} have generally larger off-axis angle is corroborated by the distribution of offsets between \xb sources and their optical NDWFS counterparts as defined by \citet{brand06}. In the right panel of Figure \ref{afig:xb_offset} it is clear that sources missing in CDWFS show larger-than-average offsets to their matched optical counterparts, likely a combination of large positional errors (deriving from large PSF sizes and low X-ray counts) and faint optical counterparts. Indeed, considering the whole \citet{brand06} catalog of \xb optical counterparts, we find that $\sim 62\%$ have an entry in the AGES optical spectroscopic catalog of \citet{kochanek12}. However, when considering the subset of \xbmissed \xb sources missing from CDWFS, only $\sim 38\%$ have an entry into the AGES catalog. Once again, this confirms that these sources are likely very faint and with a low significance. 
\par Thus, unsurprisingly, even if all \xbmissed sources would have been picked up by \texttt{wavdetect}, $\sim 88\%$ of them would have been rejected by our imposed thresholds. The remainder of the sources ($\sim 12 \%$, \xbtba) that satisfy our reliability thresholds have been added to our catalog. For the newly added sources, aperture photometry was computed for each band at the position of the source in the \citet{kenter05} catalog, extracting counts from the full resolution mosaic, analogous to what was done for the other sources. As can be seen in Table \ref{tab:xb_missing}, the majority of these significant sources (40) were genuinely missed by \texttt{wavdetect}, while few of them were missed due to reliability cuts. These latter sources are few, rare cases in which the position returned by \texttt{wavdetect} is slightly offset with respect to the position in the \citet{kenter05} catalog. This tiny offset (often around one native \chandra pixel) is enough to make the source significance fluctuate across the thresholds.

\subsection{Summary of Source Detection}
After detecting sources independently in the F, S and H bands, merging the three catalogs together into a list of \beforexb sources, and after adding \xbtba significant \xb sources, the final number of unique, X-ray point sources in our catalog is \howmany.
\par Estimating the final spurious fraction of the CDWFS catalog is not trivial, since our simulations treat each band independently (we recall that we expect on average 58, 48.5, and 52 spurious sources from simulations in the F, S and H bands, respectively), and since we added some sources from a previously published catalog. In a best-case scenario in which all the spurious sources of the F band are made up by the spurious sources of the S and H bands, which are instead detected independently, and none of the added \xb sources is spurious, we would end up with an estimated spurious fraction of $100.5/6891 = 1.5\%$. In a worst-case scenario in which the spurious sources in the F, S, and H bands are all different sources, and 1\% of the added \xb sources are also spurious, we would end up with a spurious fraction of $159/6891 = 2.3\%$. We estimated the most likely true spurious fraction by computing how many F-band spurious sources could be also detected in the S and H bands, by splitting the F-band counts into the S and H bands assuming $\Gamma=1.4$ and through Poisson statistics. On average, $14\%$ and $41\%$ of the F-band spurious sources would satisfy our reliability thresholds in the S and H bands, respectively. Furthermore, we believe the \xbtba \xb sources that satisfy our reliability thresholds to be real. Hence, the total number of spurious sources in the final catalog is expected to be $\sim127/6891 = 1.8\%$.
\par Along with the list of the excluded \xbbelowthresh \xb sources (\xbmissed$-$ \xbtba) which do not satisfy our thresholds, we release also the full electronic CDWFS catalog. An extract of the first ten sources of both catalogs can be found in Appendix \ref{asec:xbootes} (Table \ref{atab:xb_excluded} and Table \ref{atab:main_catalog}, respectively), while the full lists are available online as electronic catalogs.
\par The final detailed breakdown of the number of X-ray sources detected in each combination of the three bands is reported in Table \ref{tab:fsh}.

\begin{deluxetable}{ccc}

\tablecaption{Breakdown of the number of sources detected in each combination of bands. \label{tab:fsh}}

\tablehead{\colhead{Combination} & \colhead{Number} & \colhead{Fraction} } 

\startdata
FSH &  2498 &  36.3\% \\
FS &  2354 &  34.2\% \\
FH &  720 &  10.4\% \\
SH &  0 &  0\% \\
F &  842 &  12.2\% \\
S &  386 &  5.6\% \\
H &  91 &  1.3\% \\
\midrule
Total F & 6414 & 93\% \\
Total S & 5238 & 76\% \\
Total H & 3309 & 48\% \\
\enddata


\end{deluxetable}

\section{Number counts} \label{sec:ncts}

A typical check on the quality of an X-ray-selected catalog is to derive the number counts distribution (i.e. the number of sources per unit flux and area) and the integral of that distribution  (the \lognlogs curve). Of course, different biases have to be taken into account to derive accurate number counts. For example, the observed fluxes, measured through aperture photometry, are subject to Eddington bias (as shown also by our simulations, see Figure \ref{fig:fin-fout}). Moreover, the sensitivity of the survey has to be taken into account when correcting for missing sources at the faintest fluxes.
\par In general, number counts can be derived simply correcting the flux histogram for the incompleteness of the survey, and then integrating the number of sources per square degree and per flux bin to recover the \lognlogs. This first method, which we will refer to as `Standard', is straightforward, but has the disadvantage of ignoring Eddington bias and the fraction of spurious sources.
\par A second method is more complicated (we will refer to it as `Non Standard') but allows to correct for Eddington bias and to naturally take into account spurious sources. This method, extensively described in \citet{georgakakis08}, considers all of the sources significantly detected in a given band. Each source is detected with $\mathcal{N}$ total counts and $B$ expected background counts, within an exposure time $t_{\rm Exp}$. To take into account the uncertainty on each measured flux, a flux probability density function (PDF) can be defined:
\begin{equation}
    P(\mathcal{N},T) = \frac{T^\mathcal{N} e^{-T}}{\mathcal{N}!},
\end{equation}
where this function gives the probability to observe $\mathcal{N}$ counts given $T$, where $T$ can be explicitly written (similarly to what done in \S \ref{sec:sensitivity}) as a function of flux $s$ as $T(s) = s\mathcal{C}f_{\rm PSF}t_{\rm Exp} + B$, where $\mathcal{C}$ is the energy conversion factor from flux to count rate, and $f_{\rm PSF}$ is the encircled energy fraction of the PSF (0.9 in our case).
As discussed by \citet{georgakakis08}, if a source is detected with $\mathcal{N}$ counts, its flux PDF must also be weighted by an underlying \dnds. This factor is needed to account for the fact that there are more faint sources that could fluctuate up to higher fluxes than bright sources that fluctuate down, effectively including Eddington bias into the calculation. 
\par Since the exact shape of the underlying \dnds is what we are looking for, we applied a maximum likelihood method to obtain the best fit \dnds. First, we defined a range of fluxes $s$ and computed for each source this quantity:

\begin{equation}
  P_i =  \frac{\int P(\mathcal{N} ,T(s))\frac{dN}{ds} ds}{\int A(s) \frac{dN}{ds} ds}, 
\end{equation}
where $A(s)$ is the sensitivity curve as derived in Section \S\ref{sec:sensitivity}.
Then, the likelihood of the whole set of sources is given by $\mathcal{L} = \prod\limits_{i} P_{i} = \sum \limits_{i} \log{P_i}$. This procedure returns the best-fit parameters of the \dnds, which can be used to compute the observed number counts. Since the PDFs are defined over a large range of fluxes, extending fainter than the nominal flux limit of the survey, this method allows also to extrapolate to fluxes that are formally inaccessible to the survey.

\subsection{Simulated Number Counts}

We first explored whether we could accurately recover the input \lognlogs of the simulations. 
\par
\begin{figure*}
    \centering
    \includegraphics[width=\textwidth]{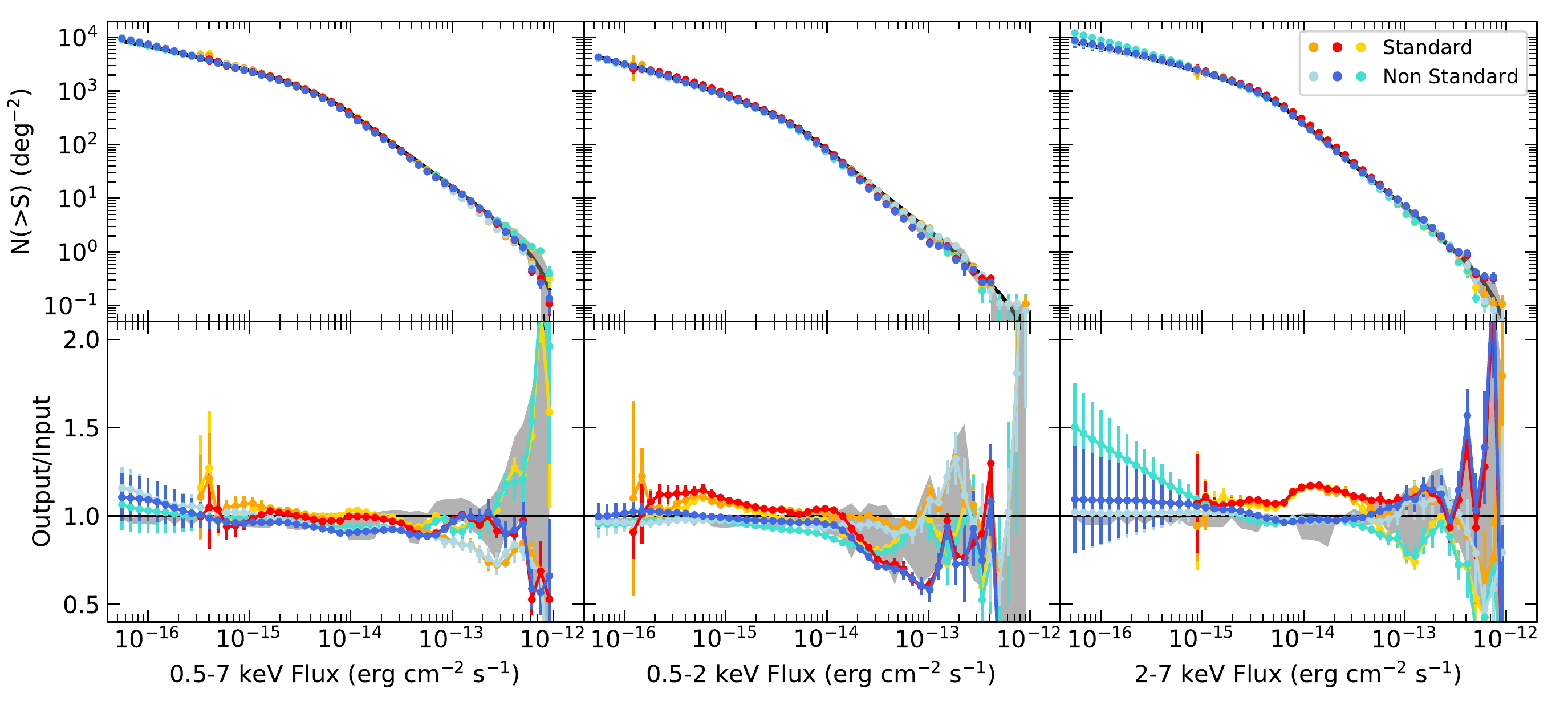}
    \caption{Performance of our measurement of the \lognlogs on simulations. From left to right panels, we show the broad, soft, and hard bands, respectively. The upper panels show the recovered \lognlogs with points and the input one with a solid black line, while the lower panels show the ratio between the points and the input \lognlogs. The sets of warm and cold colors refer to the `Standard' and `Non Standard' methods, respectively, while in each set, the color specifies to a set of ten simulations.}
    \label{fig:sim_lognlogs}
\end{figure*}

 We applied both methods and compared their robustness in recovering the input \lognlogs in our simulations. As shown in Figure \ref{fig:sim_lognlogs}, both methods are generally able to recover the input \lognlogs.
As expected, the `Non-Standard' method (the set of colder colors in Figure \ref{fig:sim_lognlogs}) better recovers the input shape. It is worth noting that the lower panels of Figure \ref{fig:sim_lognlogs} show the ratio between output/input \lognlogs where the input \lognlogs corresponds to the analytical shape of the curve, and not the actual Poissonian realization effectively used for the different sets of simulations. The spread of the measured number counts at bright fluxes is due to the Poisson fluctuations introduced by this process, depicted using gray shading in Figure \ref{fig:sim_lognlogs}.
\par As can be seen in Figure \ref{fig:sim_lognlogs}, in the hard band one simulation set overshoots the faint end of the input \lognlogs by a factor $\lesssim 1.5$. However, this happens only in the hard band, in one set out of three simulations, and only when extrapolating to fluxes below the flux limit (which is depicted in Figure \ref{fig:sim_lognlogs} by the warm colors, as the `Standard' method extends over the range of actual fluxes of the detected sources). In summary, we are confident that both methods fairly recover the underlying distribution of the number counts, and the `Non-Standard' method is more accurate and reliable over the range of fluxes covered by our sources.

\subsection{Number Counts and Resolved Fraction of the CXB}

\begin{deluxetable}{cccccc}

\tablecaption{Best fit parameters for the differential number counts for the \bootes field. \label{tab:lgnlgs}}

\tablehead{\colhead{Band} & \colhead{Sources} & \colhead{K$_{14}$} & \colhead{$S_b$} & \colhead{$\beta_1$} & \colhead{$\beta_2$}} 

\startdata
F & 6413 &  $335 \pm 2$ &  $22 \pm 3$ & $-1.78 \pm 0.03$  & $-2.59 \pm 0.08$ \\
S & 5237 & $141 \pm 2$ &  $8.1 \pm 1.2$ & $-1.68 \pm 0.03$  & $-2.56 \pm 0.09$ \\
H & 3308 & $274 \pm 1$ &  $19 \pm 5$ & $-2.03 \pm 0.06$  & $-2.76 \pm 0.10$ \\
\enddata

\tablecomments{The second column labels the number of sources used to derive the number counts. K$_{14}$ is the normalization in units of $10^{14}$ deg$^{-2}$ (\fluxcgs)$^{-1}$, while the break flux $S_b$ is in units of $10^{-15}$ \fluxcgs.}

\end{deluxetable}

\begin{figure*}
    \centering
    \includegraphics[width=\textwidth]{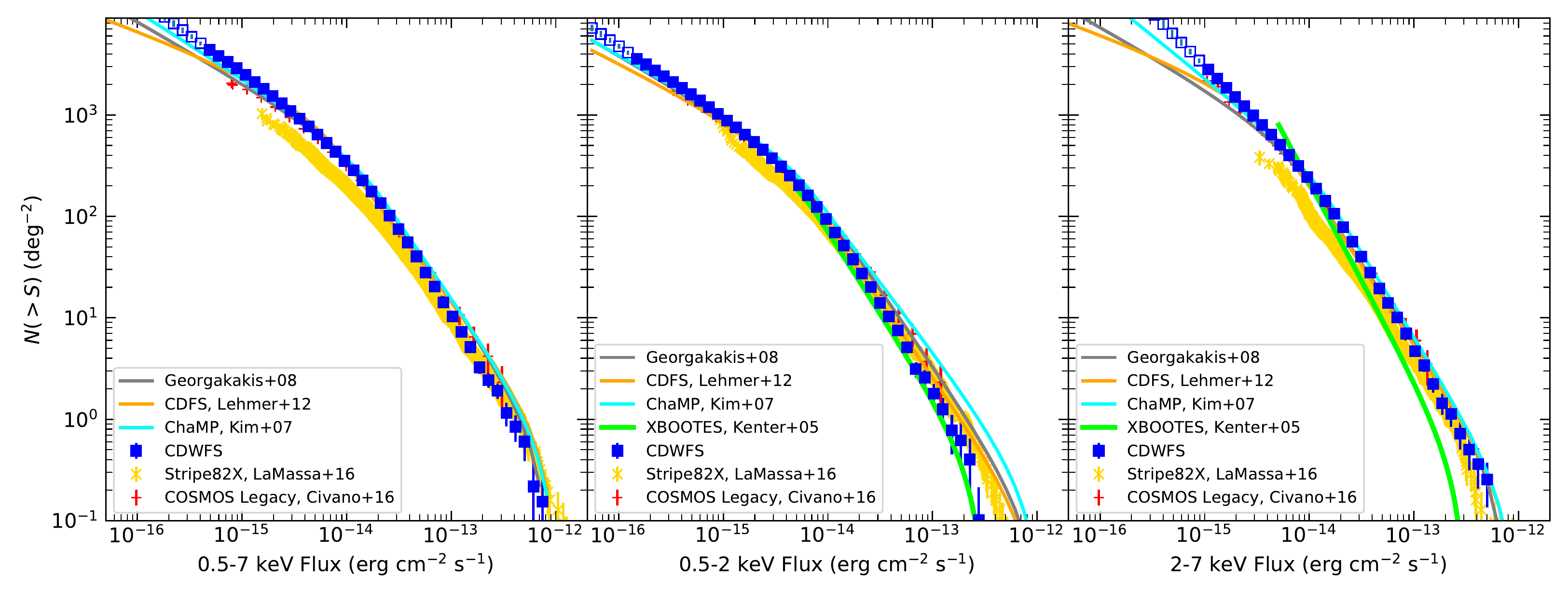}
    \caption{Measurement of the \lognlogs on data. From left to right panels, we show the broad, soft, and hard bands, respectively. Only the \lognlogs obtained from the `Non-Standard' method are shown with blue squares, with empty symbols when extrapolating below the flux limit of the survey. The results are fully consistent with the \lognlogs obtained with the `Standard' method (apart from the missing correction for Eddington bias), not shown here for clarity. As a comparison, the best-fit \lognlogs presented in \citet{georgakakis08} for a collection of \chandra surveys, \citet{lehmer12} for CDFS, and \citet{kim07} for the ChaMP survey are shown with colored lines. In the soft and hard band panels also the best fit \lognlogs relations from \xb \citep{kenter05} are shown as solid green lines. While our results are fully consistent with the \xb best fit \lognlogs in the soft band, the hard band parameters reported by \citet{kenter05} were for a single power law (i.e., not a broken power law) and have large associated uncertainties.
    Finally, the measured \lognlogs from \chandra COSMOS Legacy \citep{civano16} and Stripe82X \citep{lamassa16} are shown with red and yellow symbols, respectively. The \bootes field \lognlogs is broadly consistent with other fields in the broad and soft bands, while in the hard band there is a significant overshooting at faint fluxes, presumably due to the overdensity of sources in the deepest areas of the field, as already noted by \citet{wang04}.}
    \label{fig:dat_lognlogs}
\end{figure*}

Motivated by the previous analysis, we applied both methods to our X-ray catalog to derive robust number counts, exploiting the large number of sources detected in each band. As reported in the second column of Table \ref{tab:lgnlgs}, all sources significantly detected in each band (excluding the brightest source in the field which, as stated previously, is a star) were used in the following computation. The differential number counts were fit with a broken power law of the form

\begin{equation}
   dN/dS = \begin{cases}
               K(S_x/S_{\rm ref})^{\beta_1}, & S_x \leq S_b\\
               K(S_b/S_{\rm ref})^{\beta_1-\beta_2}(S_x/S_{\rm ref})^{\beta_2}, & S_x > S_b
            \end{cases}
\end{equation}

\noindent where $S_{\rm ref} = 10^{-14}$ \fluxcgs and $S_b$ is the break flux; the results are shown in Table \ref{tab:lgnlgs}. The uncertainties were estimated through bootstrapping. We note that in this analysis we did not separate AGNs from galaxies and stars. However, we expect galaxies to have a negligible impact on the number counts at the relatively bright fluxes probed here, with stars even less significant \citep[see][]{lehmer12}. Having computed the differential number counts, we integrated them to obtain \lognlogs, which is shown in Figure \ref{fig:dat_lognlogs} for the `Non-Standard' method. Note that the results using the `Standard' method are fully consistent with those in the Figure apart from some fluctuations, due to the different treatment of Eddington bias.

\par Figure \ref{fig:dat_lognlogs} also shows the number counts from other \chandra surveys\footnote{When the energy bands did not match exactly the ones we employed, we converted the fluxes assuming $\Gamma=1.4$.}. Since CDWFS updates the previous \xb survey, it is particularly informative to consider the best fit \lognlogs from \citet{kenter05}, in the soft and hard bands (the solid green line in Figure \ref{fig:dat_lognlogs}). In the soft band, \citet{kenter05} fit their number counts with a broken power law; our results are consistent with these results, and we can better constrain the parameters of the fit. On the other hand, \citet{kenter05} fit the hard number counts with a single power law and obtained a significantly steeper slope, while the results we obtain for CDWFS are in much better agreement with other \chandra surveys, at least at the bright end of the distribution. In particular, the faint end of the hard band \lognlogs appears to be steeper than the other measurements even before extrapolating. This higher density is found also with \lognlogs computed using the `Standard' method. \citet{wang04} also noted this higher density, and attributed it to an overdensity of faint hard X-ray sources in the deepest region of the field \citep[i.e., the LaLa survey;][]{wang04}.
\par The best-fit parameters of the differential number counts can also be used to infer the amount of resolved fraction of the CXB: this is trivially done computing the integral 
\begin{equation}
    \int_{S_{\rm min}}^{S_{\rm max}} \frac{dN}{dS}SdS,
\end{equation}
where $S_{\rm min}$ and $S_{\rm max}$ are the minimum and maximum flux of the detected sources. The result of this integral gives the total X-ray flux per square degree resolved into single sources. Thus, the fraction of resolved CXB is obtained by comparing the resolved flux to the total CXB intensity in a given energy band. Converting the values reported by \citet{hickoxmarkevitch06} to match our energy bands assuming $\Gamma = 1.4$, we obtained a resolved fraction of \softcxb and \hardcxb for the soft and hard bands, respectively. The uncertainties were estimated assuming a normal PDF for the parameters of the \dnds, and solving the integral 5000 times, each time randomly picking the parameters according to their PDFs. The uncertainty on the total intensity of the CXB as reported by \citet{hickoxmarkevitch06} was propagated to obtain the error on the resolved fraction. Since in this paper we detected only point sources, the resolved flux does not take into account the contribution from extended sources (such as galaxy clusters or groups), which is expected to be larger in the soft band. The resolved fraction of the CXB is consistent within the uncertainties with that reported by \citet{kim07b} for the ChaMP survey in both the S and H bands, at comparable depth and total area with CDWFS; also, in the H band, with that reported by \citet{hickoxmarkevitch06}.

\section{Optical-IR counterparts} \label{sec:opt}

\begin{figure}
    \centering
    \includegraphics[width=0.47\textwidth]{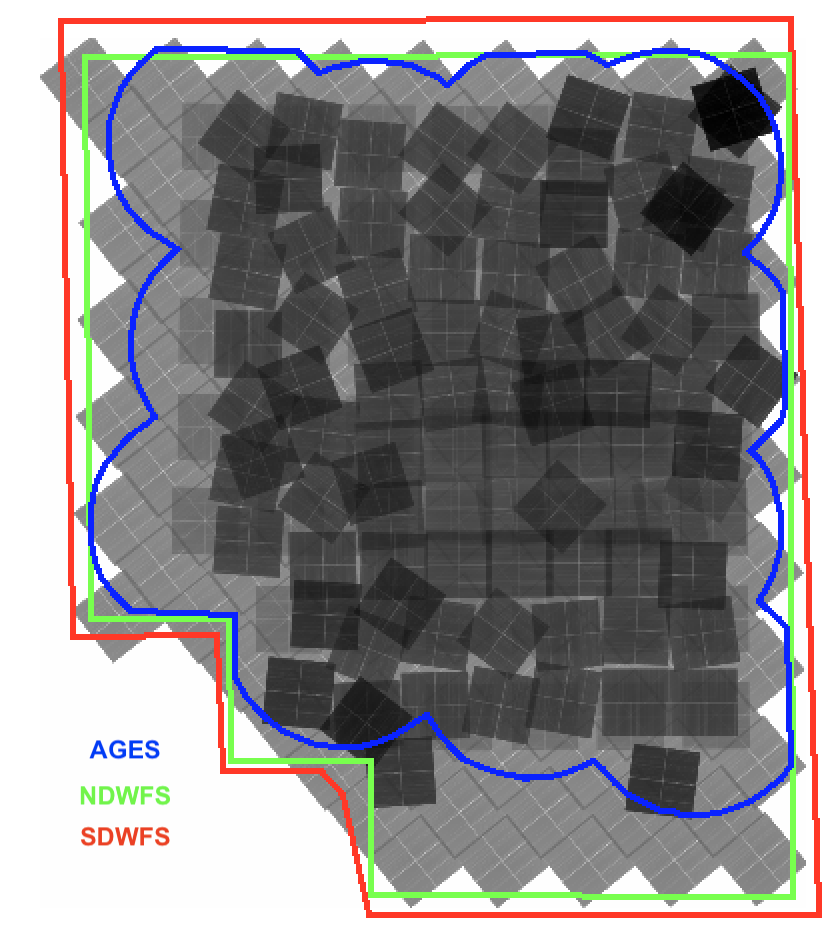}
    \caption{The multi-wavelength coverage of the \bootes field considered in this work. The green region labels the photometric coverage provided by the optical NDWFS \citep{jannuzi&dey99}, while the red one labels the MIR SDWFS coverage \citep{ashby09}. Optical spectroscopy is provided by the AGES survey \citep{kochanek12} and is labeled in blue.}
    \label{fig:mwcov}
\end{figure}

The \bootes field has rich multi-wavelength coverage, from the radio to the X-rays, which we briefly summarize here. In the radio, many frequencies are covered: including LOFAR LBA \citep[50 MHz;][]{vanweeren14} and HBA \citep[150 MHz;][]{retanamontenegro18}, as well as GMRT \citep[153 MHz;][]{williams13}, VLA \citep[325 MHz;][]{coppejans15}, and 1.4 GHz WSRT \citep{devries02}. Far-IR data are available from the \textit{Herschel} Multi-tiered Extragalactic Survey \citep[HerMES;][]{oliver12}, providing data from both SPIRE at 250, 350, and 500 $\mu$m, and PACS at 110 and 170$\mu$m.
\par Mid-IR coverage is available from both the \textit{Spitzer} MIPS AGN and Galaxy Evolution Survey \citep[MAGES;][]{jannuzi10} and the \textit{Spitzer} Deep Wide-Field survey \citep[SDWFS;][]{ashby09,kozlowski10,kozlowski16}, complementing the earlier \textit{Spitzer} IRAC Shallow Survey \citep{eisenhardt04} in the four IRAC channels. Near-IR data in the $J$, $H$ and $K_s$ bands are available from the NEWFIRM Infrared \bootes Imaging Survey \citep[IBIS;][]{gonzalez10}. $Y$-band data are available from LBT \citep{bian13}. Optical photometry is provided by NDWFS in the $B_W$, $R$, and $I$ bands \citep{jannuzi&dey99}, by LBT in the $U_{\rm spec}$ band \citep{bian13}, and by z\bootes in the $z$ band \citep{cool07}. Multi-band photometry for $I$-band detected sources is also available \citep{brown07}. In addition, optical spectroscopy is available from the AGN and Galaxy Evolution Survey \citep[AGES;][]{kochanek12}. UV coverage is provided by \textit{GALEX}, and extensive X-ray coverage from \chandra is available from our work (CDWFS) and from the previous \xb survey \citep{murray05}.
\par Following a standard approach adopted in X-ray surveys, we matched our X-ray catalog to optical and mid infrared (MIR) data adopting the maximum likelihood ratio technique \citep[e.g.,][]{brusa05,brusa07,brusa10,marchesi16}. In particular, we matched with the full $I$-band NDWFS ($3\sigma$ depth of 22.9 Vega magnitudes) and [3.6]-selected SDWFS ($5\sigma$ depth of 19.77 Vega magnitudes) catalogs, whose spatial coverages are shown in Figure \ref{fig:mwcov}. We used NWAY \citep{salvato18} to simultaneously match our X-ray catalog to the $I$-band and [3.6]-band catalogs.
\par Each X-ray source in our catalog was associated to its most likely optical/IR counterpart taking into account both the distance and the magnitude of the candidate counterpart. We considered all the counterparts within $10\arcsec$ of each CDWFS source, and used a much stricter radius of $1\arcsec$ to build the magnitude histogram of the true counterparts, as done, e.g., by \citet{marchesi16}. While we used the computed positional error of our X-ray sources, a positional error of $0\farcs1$ for the $I$-band catalog and $0\farcs5$ for the [3.6]-band catalog were assumed.
\par
\begin{figure*}
    \centering
    \includegraphics[width=0.47\textwidth]{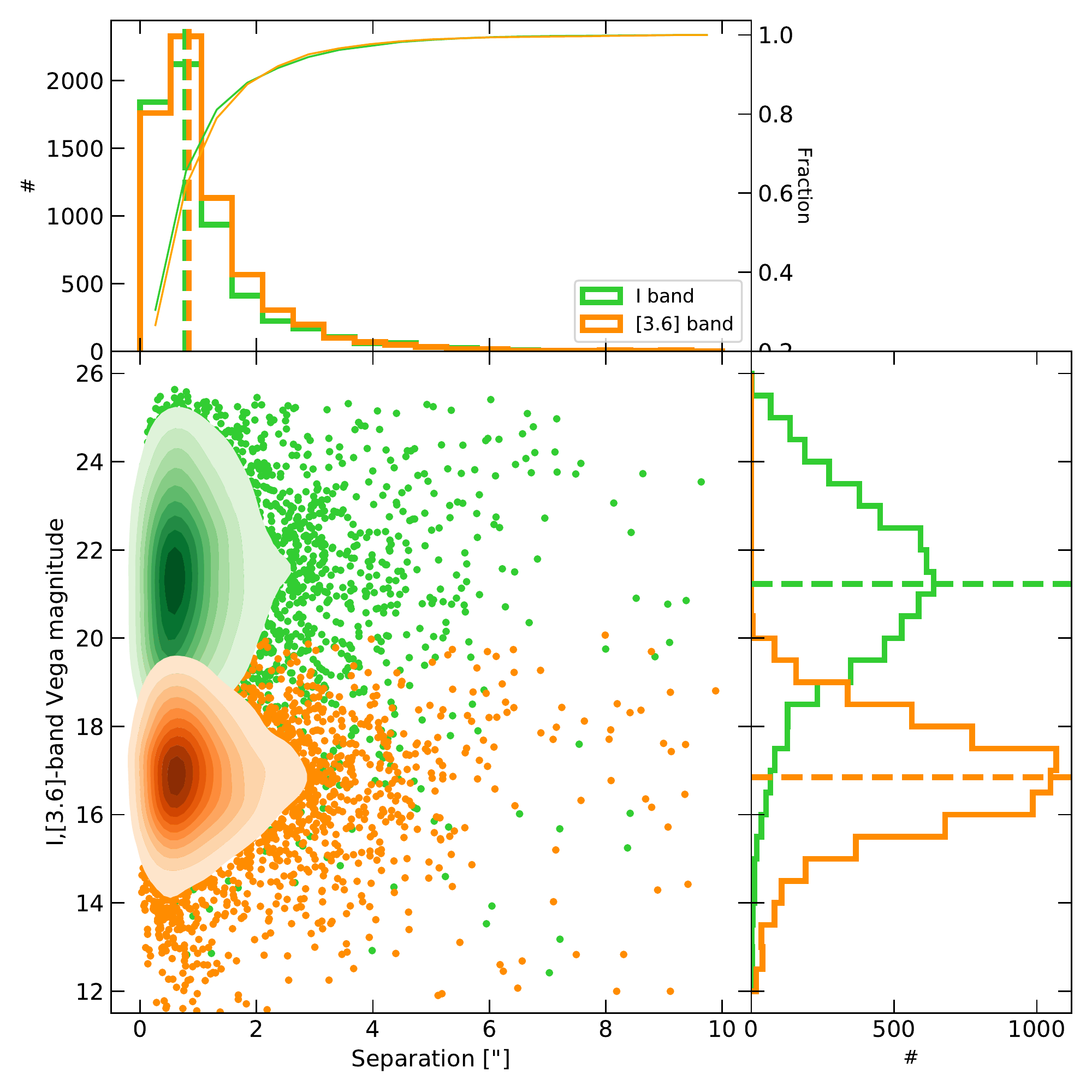}
    \includegraphics[width=0.47\textwidth]{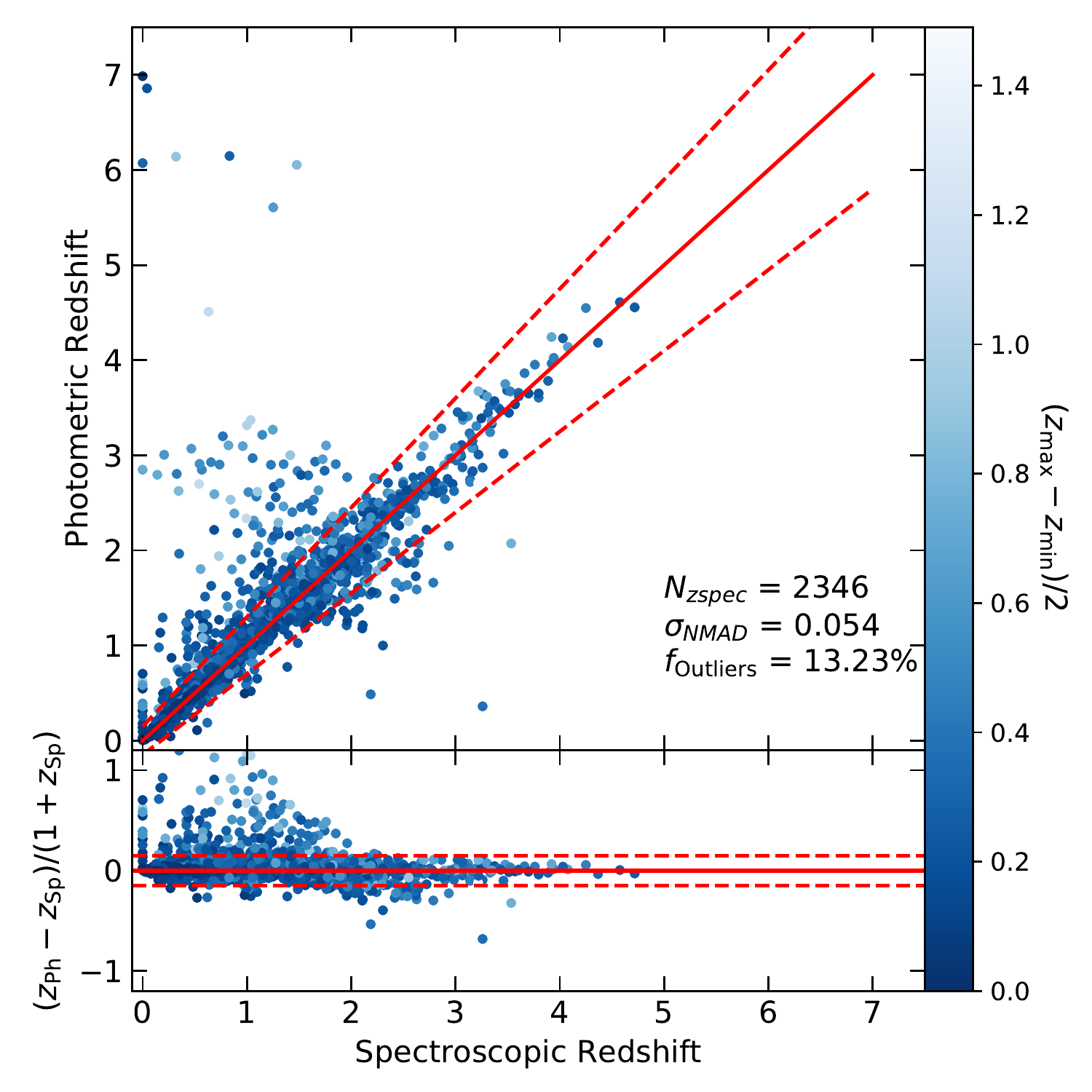}
    \caption{\textbf{Left.} The central panel shows the separation-magnitude plane for all the CDWFS sources with an $I$-band and/or [3.6]-band counterpart, shown as green and orange dots, respectively. The overlaid contours qualitatively show the density gradient of sources in the plane. The top panel shows the separation distributions. The medians of the distributions are labeled by vertical dashed lines and are both $\sim 0\farcs8$. The cumulative distributions are also shown as thin lines, and show that $90\%$ of counterparts are matched within 2\arcsec. The right panel shows the magnitude distributions for the two samples, with the medians of the distributions labeled as dashed lines. \textbf{Right.} Comparison between photometric and spectroscopic redshifts for the 2346 sources with a valid spectroscopic redshift entry. Points are color-coded by their approximate uncertainty. The solid red line marks the 1:1 relation, while the dashed red lines limit the region of $\Delta z/(1+z_{\rm sp}) = \pm 0.15$. Sources outside this region are marked as outliers, and represent $\sim 13\%$ of the total. This figure highlights the very good agreement between photometric and spectroscopic redshifts in the \bootes field.}
    \label{fig:mwprops}
\end{figure*}

 The matched catalog returned by NWAY was cleaned retaining only the most likely (if any) counterparts (i.e. those with \texttt{match\_flag=1}). The left panel of Figure \ref{fig:mwprops} shows the magnitude distributions of multiwavelength counterparts, and their distances from our X-ray sources. The large majority ($\sim 90\%$) of the counterparts are matched within $2\arcsec$. Each X-ray source has a probability of being associated with its correct counterpart (the parameter \texttt{p\_any}), and we estimated (by randomly shuffling the X-ray catalog and re-matching it with the same maximum likelihood ratio procedure) that a cut of \texttt{p\_any} $>$ \panycut is required to ensure fewer than 10\% spurious associations\footnote{To have fewer than (5\%, 3\%, 1\%) spurious associations, \texttt{p\_any} cuts of (0.68, 0.85, 0.96) should be used.}. Out of \howmany X-ray sources, 6843 (5852) have at least one counterpart in the $I$-band and/or [3.6]-band catalogs with \texttt{p\_any} $> 0$ (\texttt{p\_any} $>$ \panycut).

\subsection{Redshifts and X-ray to Optical Properties}

\begin{deluxetable}{lcc}

\tablecaption{Results of the multi-wavelength match of CDWFS catalog with the NDWFS and SDWFS catalogs. \label{tab:mwnumbers}}

\tablehead{\colhead{} & \colhead{\texttt{p\_any} $>$ 0} & \colhead{\texttt{p\_any} $>$ \panycut}} 
\startdata
X-ray sources &  \howmany &  5852  \\
Count. in at least one band &  6843 &  5852 \\
Count. in both bands & 5802  &  5393  \\
Available redshifts &  6449 &  5661  \\
Best redshift spectroscopic &  2346 &  2287  \\
Best redshift photometric  &  4103 &  3374  \\
\enddata


\end{deluxetable}

To obtain spectroscopic redshifts, we matched the CDWFS catalog with the AGES catalog \citep{kochanek12}, using the NDWFS $I$-band coordinates (or the SDWFS ones, for the 810 sources with IR-only counterparts) and adopting a matching radius of $0\farcs5$. We shifted the whole AGES catalog in order to correct its astrometry with respect to \text{Gaia} (see Appendix \ref{asec:astrometry}). In total, we had 2346 valid spectroscopic redshifts. Then, we exploited the recent release of hybrid photometric redshifts \citep[template $+$ machine learning; see][]{duncan18a,duncan18b} for the upcoming data release of the LOFAR Two-metre Sky Survey Deep Fields (Duncan et al. 2020, submitted). Analogous to that done for the spectroscopic redshifts, we matched our CDWFS sources with a matching radius of 0\farcs5, and choosing the closest counterpart when multiple $I$-band entries were found within this distance. We ended up with a total of 6447 photometric redshifts.
The overall quality of the photometric redshifts compared to the spectroscopic ones is shown in Figure \ref{fig:mwprops} and it is generally very good. A handful of outliers show a photometric redshift higher than six, but are all actually associated to much lower spectroscopic redshifts. We thus suggest caution when focusing on the $30$ CDWFS sources with $z_{\rm Phot} \gtrsim 6$. The normalized median absolute deviation, $\sigma_{\rm NMAD} = 1.48 \times \text{median}(\lvert z_{\rm ph} - z_{sp}\rvert/(1+z_{\rm sp}))$, is $0.054$, while the fraction of outliers, defined as the fraction of sources where $ \lvert z_{\rm ph} - z_{sp}\rvert/(1+z_{\rm sp}) > 0.15$, is $13.23\%$. These already excellent values, will soon be likely further improved by the upcoming Subaru HSC data coverage of the \bootes field. Table \ref{tab:mwnumbers} summarizes the numbers of counterparts and redshifts for the whole X-ray catalog with \texttt{p\_any} $>$ 0 and for the subset for which \texttt{p\_any} $>$ \panycut.

\subsection{Hardness Ratio and Obscuration}

Once the redshift of a given X-ray source is known, a rest-frame luminosity can be computed from the observed flux. X-rays, although very penetrating, suffer an absorption bias depending on the amount of gas column density along the line of sight (usually denoted with $N_{\rm H}$). This absorption effect is band-dependent, and mostly affects soft X-rays. To derive statistically reliable absorption-corrected, rest frame luminosities, we first estimated the gas column density $N_{\rm H}$ for any given source. We used a standard method, which estimates the column density using the combination of the hardness ratio (HR, defined as HR = (H-S)/(H+S), where H and S are the hard and soft counts, respectively) and the redshift information. The HR was computed for each X-ray source using the Bayesian Estimator for Hardness Ratio \citep[BEHR,][]{park06}, and taking into account the slightly larger area from which the hard counts have been extracted during the aperture photometry step, due to the larger PSF size in the H band.
\par The HR encodes the observed spectral shape of the source; since the absorption bias is band-dependent, the redshift has to be known to properly estimate the intrinsic column density at the location of the source. We therefore created a grid of theoretical curves of HR as a function of redshift for a large range of column density, using XSPEC \citep{arnaud96} and a simple power law with photon index $\Gamma=1.8$, consistent with the average, absorption-corrected value of the whole AGN population \citep[e.g.,][]{ricci17}. We modified the assumed power law to account for absorption by a fixed Galactic column density $N_{\rm H, Gal} = 1.04 \times 10^{20}$ cm$^{-2}$ \citep{kalberla05}, estimated approximately at the center of the \bootes field, and for absorption by an additional amount of column density in the range log($N_{\rm H}$/cm$^{-2}$) $= [20,25]$ with a step of 0.02 dex\footnote{It should be noted that theoretical $\rm{HR}(z)$ functions were obtained with model fluxes, which were then converted to count rates using the appropriate exposure-weighted energy conversion factor on a source-by-source basis. Here, we are implicitly ignoring the difference in exposure time between the S and H bands, so that an HR computed with count rates corresponds to one computed with observed counts. This simplification is justified because $95\%$ of our sample has an H-to-S exposure ratio in the range $0.95$--$1.15$.}. 
\par Once the correlated parameters HR and $z$ were known for a given source, the closest $N_{\rm H}$-dependent theoretical curve in our grid was used to estimate the column density that, for a typical power law of $\Gamma=1.8$ at redshift $z$, resulted in the observed HR (and ultimately on the observed S and H-band counts). 
\par It is worth stressing that the spectral shape of a local heavily obscured AGN is much more complex than what is defined by a simple obscured power-law. First, the assumed simple model does not include the effect of Compton scattering, which becomes important at the highest column densities, close to the Compton-thick threshold (i.e., $N_{\rm H} \gtrsim 10^{24}$ cm$^{-2}$); ignoring it leads to an underestimation of the intrinsic luminosity for such highly obscured AGNs \citep[e.g.,][]{li19}. Moreover, the possible presence of a reflection component would harden the spectrum, resulting in an overestimation of the obscuration \citep[e.g.,][]{wilkes13}. In addition, prominent soft emission, likely arising from photons that have scattered off clouds on a scale much larger than the nuclear one (where most of the obscuring material lies) dominates the soft X-ray spectrum of heavily obscured AGNs at $E \lesssim 2-3$ keV \citep{bianchiguainazzi07,ricci17}. This implies that, for high column densities approaching the Compton-thick level and for low-to-intermediate redshifts, the HR estimated from a more realistic model can be much softer than what would be expected from a simple obscured power-law. However, if we try to incorporate a more physical model when creating a grid of theoretical HR$(z)$ tracks, many tracks end up crossing each other, invalidating our method of using the HR-$z$ plane to uniquely estimate individual source properties. The effect of such a simplification on estimates of column density is that a fraction of AGNs appearing as unobscured and faint, will, in reality, be obscured and intrinsically more luminous \citep{lambrides20}. A careful assessment of the obscuration state of our sample is outside the scope of the present paper, and requires a combination of multi-wavelength diagnostics, e.g. using rest-frame MIR luminosities and X-ray spectral analysis, to disentangle genuine, faint and unobscured AGNs from luminous obscured ones.
\par
\begin{figure*}
    \centering
    \includegraphics[width=0.47\textwidth]{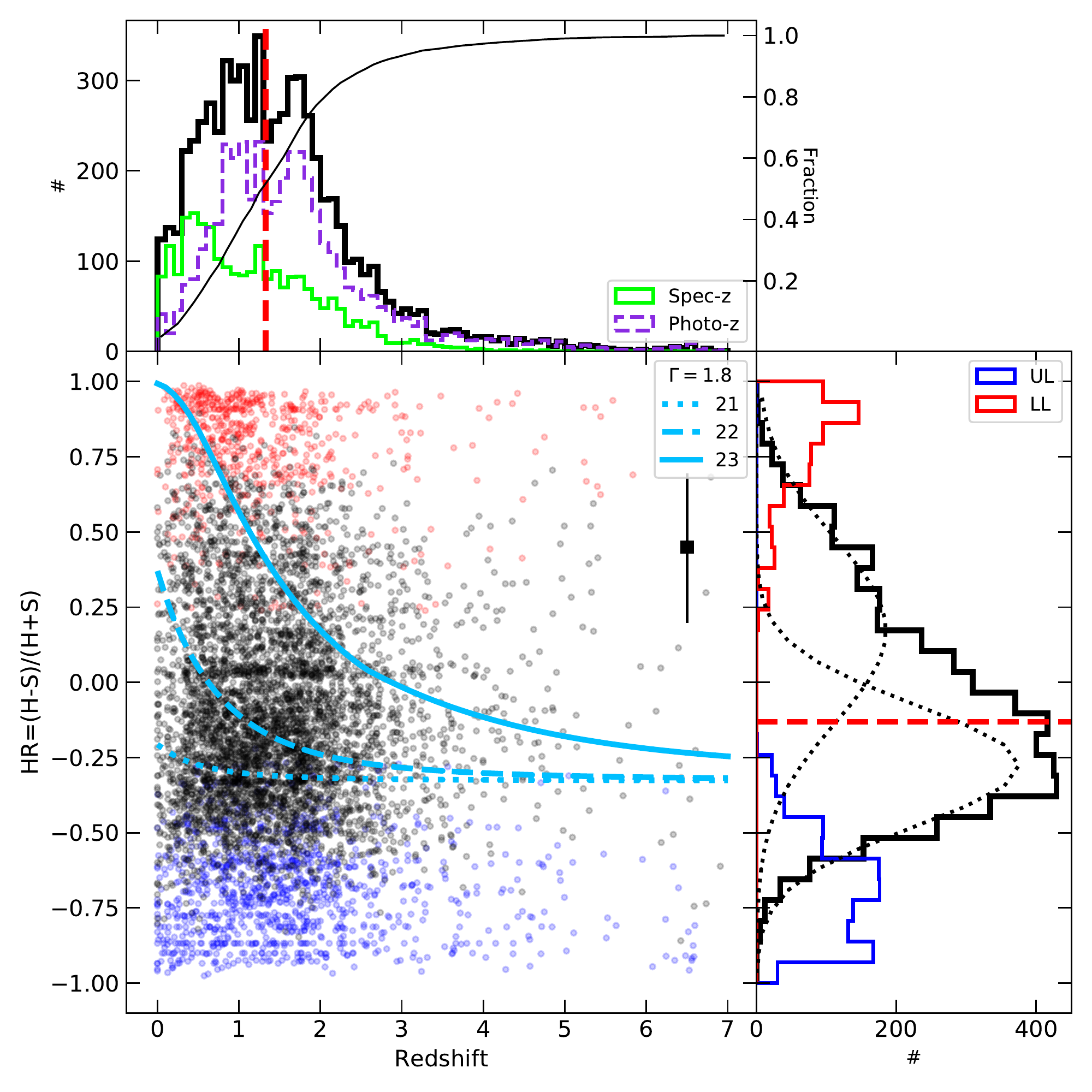}
    \includegraphics[width=0.47\textwidth]{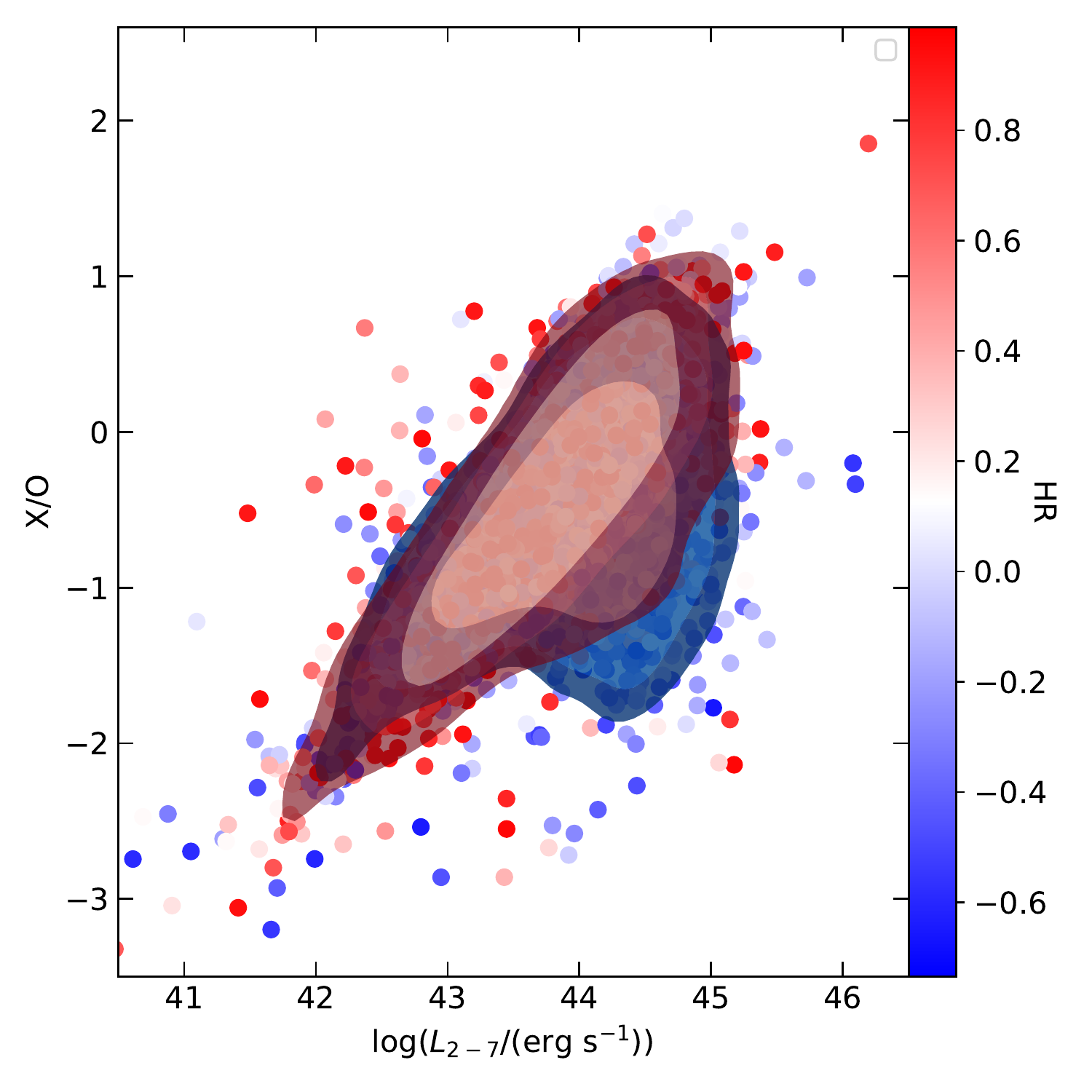}
    \caption{\textbf{Left.} The central panel shows the HR-$z$ plane for all the CDWFS sources with a redshift (i.e. 6465). The constrained HRs (i.e., the ones which do not reach the lower or upper allowed values within their $1\sigma$ uncertainties) are shown as black points, while upper limits and lower limits are shown with blue and red points, respectively. The median uncertainty on HR for the constrained sample is shown as the black point on the right. The cyan tracks show three different HR$(z)$ functions for log($N_{\rm H}$/cm$^{-2}$) $= 21,22,23$ in dotted, dashed and solid lines, respectively. The top panel shows the redshift distribution of the sample, split into photometric (dashed violet histogram) and spectroscopic (solid green histogram). The median of the distribution is labeled by a red vertical dashed line. The cumulative distribution is also shown as a thin black line. The right panel shows the HR distribution for the three samples (constrained in solid black, upper/lower limits in solid blue/red), with the median of the distribution labeled as a dashed red line. Also shown in dotted lines is the two-Gaussian decomposition of the black histogram. \textbf{Right.} X-ray-to-optical flux ratio (X/O) as a function of hard band luminosity. Sources are color-coded by their HRs, and density contours are overlaid (red contours for 68\%, 90\% and 99\% of sources with HR$>0.2$, and blue contours for the same fraction of sources with HR$<0$). Despite large scatter due to the uncertainties on each single HR, softer sources (bluer points and contours) are more spread out to lower X/O ratio at $L_x > 10^{43}$ \lumcgs, as expected for unobscured AGNs.}
    \label{fig:hrz}
\end{figure*}

 The HR-$z$ plane, together with the relative redshift and HR distributions of our sample, is shown in the left-hand panel of Figure \ref{fig:hrz}. 
As can be seen from the upper section of this panel, $\sim 90\%$ of our sample lies within $z<3$. The HR distribution of the whole sample, shown on the right-hand side of the left-hand panel of Figure \ref{fig:hrz} 
demonstrates that the median HR of the whole (HR-constrained) sample ($\overline{\rm HR}=-0.16$) is similar to the expected HR for an unobscured power law with photon index $\Gamma=1.8$, but shifted to harder (i.e. larger) HR. Indeed, the whole HR distribution looks asymmetric, and is well-fit by a double Gaussian function \citep[e.g.][]{civano16}, with ($\mu, \sigma) = (0.22,0.29)$ and $(-0.27,0.21)$, representing a mix of the unobscured and obscured populations of AGNs.
\par Since the early ages of X-ray surveys, AGNs have been known to occupy a well-defined region of the X-ray and optical flux parameter space. In particular, the ratio between the X-ray and optical flux, defined as X/O $=  \log{f_x} + m/2.5 + C$ \citep[where $C$ is a constant dependent on the photometric band considered; e.g.,][]{tananbaum79, maccacaro88} is generally in the range [-1,1]. Moreover, correlations have been found between X/O and the X-ray luminosity, mainly for obscured AGNs \citep{fiore03,marchesi16}: at any given X-ray luminosity, obscured AGNs are expected to have their UV/optical emission fainter than their unobscured counterparts, resulting in a higher X/O ratio. Hence, we can further check the broad correspondence between HR and obscuration by investigating the behavior of the X/O ratio as a function of hard band luminosity, using the H-band intrinsic luminosity and the $i$-band magnitude to compute X/O; we estimated $C=5.03$ for the NDWFS $i$-band filter, although its exact value is not important as it just shifts the whole figure to higher or lower values of X/O. In the right panel of Figure \ref{fig:hrz}, we color-code sources based on their HR and through their density contours visually demonstrate that, in general, hard sources fall in the region of the plane where obscured AGNs are expected to be. Likewise, soft sources are more spread out, in particular at $L_X > 10^{43}$ \lumcgs.

\subsection{Intrinsic Luminosity}

Having estimated $N_{\rm H}$ values using the HR-$z$ plane, we computed the correction factor $k$, defined as the ratio between the obscured and intrinsic flux $k = f_{\rm obs}/f_{\rm int}$, where $f_{\rm int}$ is estimated with an unobscured power law with $\Gamma=1.8$, following \citet{marchesi16}. Since our $N_{\rm H}$ range is limited to log($N_{\rm H}$/cm$^{-2}$) $\geq 20$, we assigned $k=1$ to those sources with log($N_{\rm H}$/cm$^{-2}$) $\leq 20$. We translated one-sigma uncertainties on the HR into uncertainties on the column density, which in turn constrained the minimum and maximum values of $k$, for the lower and higher $N_{\rm H}$ limits, respectively.
\par Once we have calculated $k$ for each band, we computed the rest-frame, intrinsic flux $f_{\rm int}^{\rm RF} = f_{\rm int}(1+z)^{\Gamma-2}$ (with $\Gamma=1.8$), and then, finally, the intrinsic luminosity at a given redshift using $L_{\rm int}^{\rm RF} = 4\pi D_{L}^2 f_{\rm int}^{\rm RF}$, where $D_L$ is the luminosity distance.
\par
\begin{figure*}
    \centering
    \includegraphics[width=0.47\textwidth]{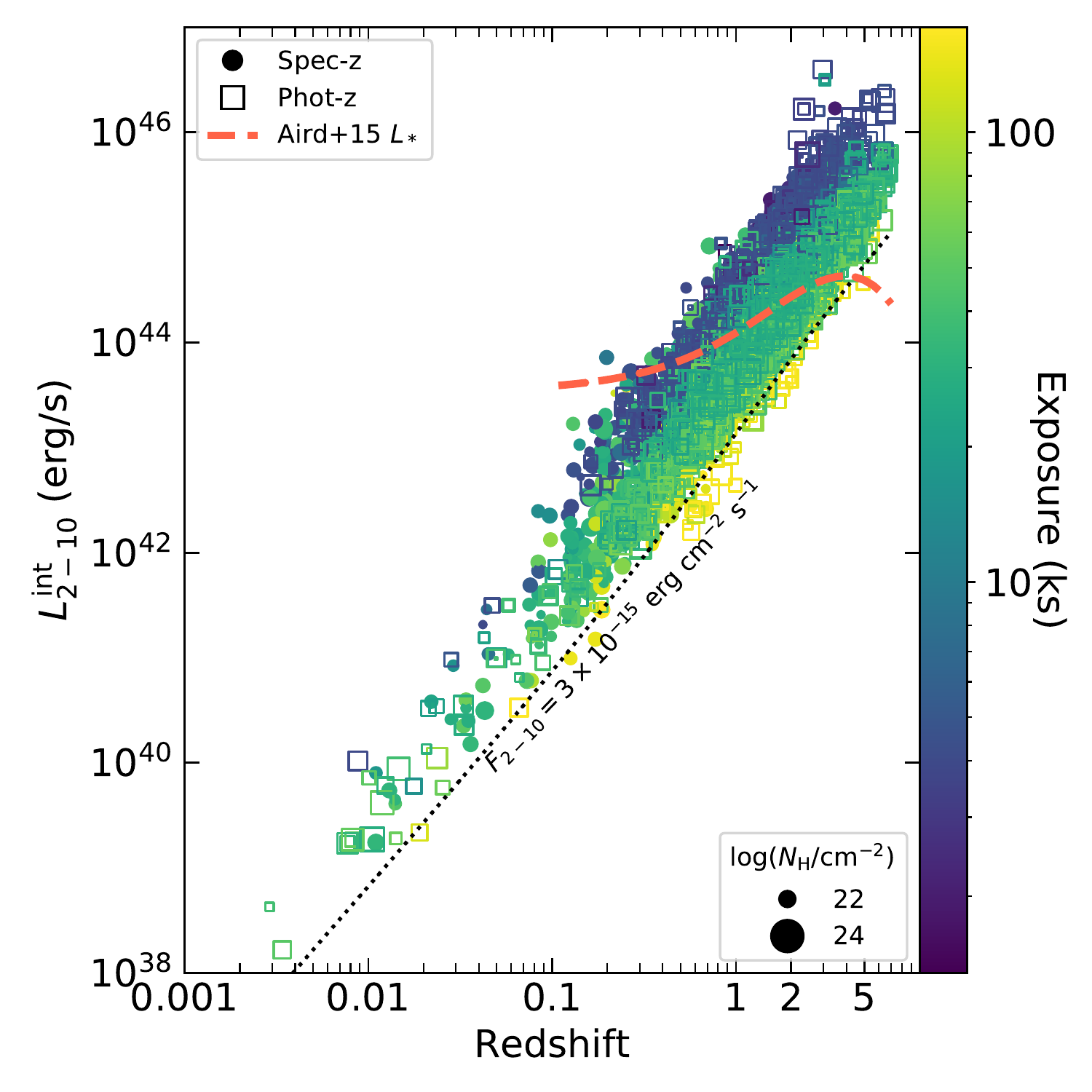}
    \includegraphics[width=0.47\textwidth]{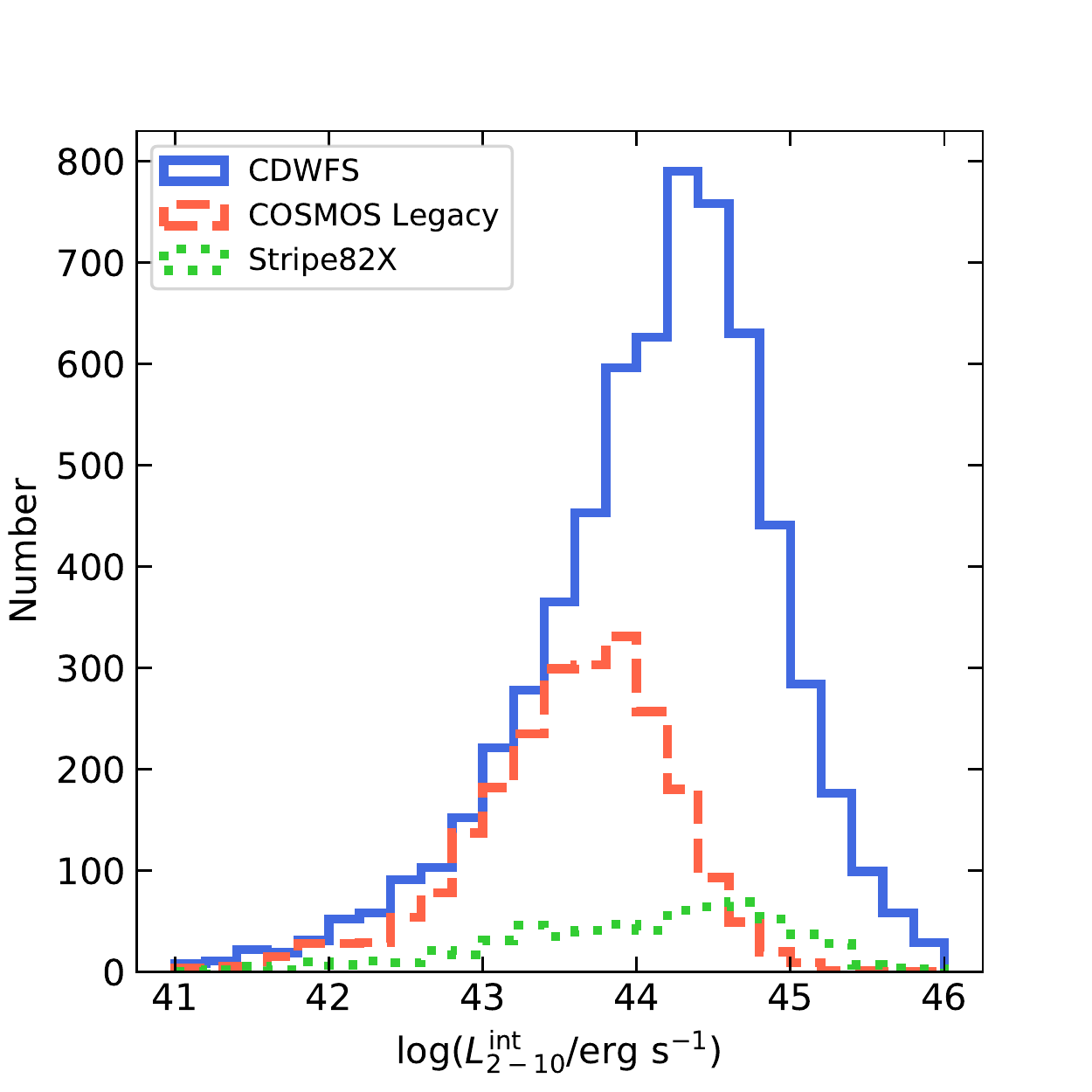}
    \caption{\textbf{Left.} Intrinsic (i.e. corrected for absorption), rest frame $2-10$ keV luminosity as a function of redshift. Datapoints with spectroscopic/photometric redshifts are labeled as filled circles/empty squares, respectively, their size is proportional to the $N_{\rm H}$ derived through the HR-$z$ plane, and are color-coded by exposure (in ks). The dotted line shows a flux of $3\times10^{-15}$ \fluxcgs, while the red dashed line labels the evolution of the knee of the luminosity function ($L_*$) from \citet{aird15}. \textbf{Right.} Intrinsic, rest frame $2-10$ keV luminosity  distribution for CDWFS (solid blue histogram), COSMOS Legacy \citep[dashed red]{civano16,marchesi16}, and Stripe82X \citep[dotted green]{lamassa16}. With its unique features, CDWFS is a much valuable addition in the array of \chandra's Extragalactic Legacy Surveys, peaking at luminosities in between the other two major wide surveys, and providing excellent statistics.}
    \label{fig:lums}
\end{figure*}

 A plot of the rest frame, intrinsic luminosity in the 2-10 keV band (converted from the 2-7 keV luminosity assuming $\Gamma=1.8$) as a function of redshift is shown in the left panel of Figure \ref{fig:lums}. 
This panel demonstrates that the range of exposure times for observations in the \bootes field allows a relatively broad range of the luminosity-redshift plane to be explored. In particular, thanks to its combination of wide area and deep exposures, the CDWFS survey has sufficient statistical power to probe both above and below the knee of the luminosity function \citep[$L_*$, as parametrized by][]{aird15} in the redshift range $z\sim 0.5$--$3$. This is crucial to probe the full distribution of mass accretion rates for the AGN population over a large redshift range. This is even more evident when the distribution of rest-frame $2$--$10$ keV luminosities for CDWFS, COSMOS Legacy \citep{civano16,marchesi16}, and Stripe82X \citep{lamassa16} are compared, as demonstrated in the right panel of Figure \ref{fig:lums}. We show the $2$--$10$ keV band luminosity since it is less affected by obscuration corrections, but the situation is the same for the soft band. The combination of the wide area and the depth accumulated through 15 years of \chandra observations makes CDWFS unique in terms of coverage.
CDWFS alone has more sources than the COSMOS Legacy and Stripe82X surveys combined, at any luminosity. Furthermore, the distribution of CDWFS luminosities bridges the gap between COSMOS Legacy and Stripe82X, adding a significant number of sources in the luminosity range log$(L/$erg s$^{-1}$) $\sim 44$--$45$. This is crucial to probe the typical \textit{and} most luminous accreting black holes at the peak epoch of AGN and galaxy co-evolution.

\section{Catalog description} \label{sec:catalog}

The final X-ray catalog, which we release in electronic form, contains \howmany sources and contains a rich amount of information, with \howmanycols columns. An extract of the first ten sources can be found in Appendix \ref{asec:xbootes} (Table \ref{atab:main_catalog}).
\par The catalog is mainly divided into four sections: 
\begin{enumerate}
    \item Columns which are exclusively related to purely X-ray-derived quantities, such as coordinates, probabilities, counts, and fluxes.
    \item Columns which start with the syntax `NDWFS\_' are related to the NDWFS $I$-band catalog used by NWAY to look for $I$-band counterparts to X-ray sources.
    \item Columns which start with the syntax `SDWFS\_' are related to the SDWFS [3.6]-band catalog used by NWAY to look for [3.6]-band counterparts to X-ray sources.
    \item Other columns are related to derived quantities, either from NWAY, or to obtained redshifts, or from deriving final absorption corrections and intrinsic luminosities.
\end{enumerate}
The breakdown of the columns of the catalog is as follows:
\newline \textit{Column 1}. ID of the X-ray source.
\newline \textit{Columns 2-3}. RA and DEC (J2000) of the X-ray source, in degrees.
\newline \textit{Column 4}. Positional error of the X-ray source, in arcsec.
\newline \textit{Column 5}. F-band Poissonian probability that the source is a background fluctuation.
\newline \textit{Column 6}. F-band exposure-weighted average $r_{90}$, the approximate circular radius encompassing 90\% of \chandra's PSF (in arcsec). This radius was used to extract aperture photometry.
\newline \textit{Column 7}. F-band total counts within a circle of radius $r_{90}$.
\newline \textit{Column 8}. F-band background counts within a circle of radius $r_{90}$, extracted from our background maps at the position of the source.
\newline \textit{Column 9}. F-band net counts within a circle of radius $r_{90}$, or $3\sigma$ upper limit if the probability of being spurious of the source is higher than the threshold.
\newline \textit{Columns 10-11}. Positive and negative errors on F-band net counts, or $-99$ if the net counts are a $3\sigma$ upper limit.
\newline \textit{Column 12}. F-band total exposure in seconds at the position of the source.
\newline \textit{Column 13}. F-band net count rate in cts/s, or $3\sigma$ upper limit if the probability of being spurious of the source is higher than the threshold.
\newline \textit{Columns 14-15}. Positive and negative errors on F-band net count rate, or $-99$ if the net count rate is a $3\sigma$ upper limit.
\newline \textit{Column 16}. F-band flux, or $3\sigma$ upper limit if the probability of being spurious of the source is higher than the threshold. The flux is obtained from the count rate (or its upper limit) applying an aperture correction and the appropriate exposure-weighted energy conversion factor assuming $\Gamma=1.4$.
\newline \textit{Columns 17-18}. Positive and negative errors on F-band flux, or $-99$ if the flux is a $3\sigma$ upper limit.
\newline \textit{Column 19}. S-band Poissonian probability that the source is a background fluctuation.
\newline \textit{Column 20}. S-band exposure-weighted average $r_{90}$, the approximate circular radius encompassing 90\% of \chandra's PSF (in arcsec). This radius was used to extract aperture photometry.
\newline \textit{Column 21}. S-band total counts within a circle of radius $r_{90}$.
\newline \textit{Column 22}. S-band background counts within a circle of radius $r_{90}$, extracted from our background maps at the position of the source.
\newline \textit{Column 23}. S-band net counts within a circle of radius $r_{90}$, or $3\sigma$ upper limit if the probability of being spurious of the source is higher than the threshold.
\newline \textit{Columns 24-25}. Positive and negative errors on S-band net counts, or $-99$ if the net counts are a $3\sigma$ upper limit.
\newline \textit{Column 26}. S-band total exposure in seconds at the position of the source.
\newline \textit{Column 27}. S-band net count rate in cts/s, or $3\sigma$ upper limit if the probability of being spurious of the source is higher than the threshold.
\newline \textit{Columns 28-29}. Positive and negative errors on S-band net count rate, or $-99$ if the net count rate is a $3\sigma$ upper limit.
\newline \textit{Column 30}. S-band flux, or $3\sigma$ upper limit if the probability of being spurious of the source is higher than the threshold. The flux is obtained from the count rate (or its upper limit) applying an aperture correction and the appropriate exposure-weighted energy conversion factor assuming $\Gamma=1.4$.
\newline \textit{Columns 31-32}. Positive and negative errors on S-band flux, or $-99$ if the flux is a $3\sigma$ upper limit.
\newline \textit{Column 33}. H-band Poissonian probability that the source is a background fluctuation.
\newline \textit{Column 34}. H-band exposure-weighted average $r_{90}$, the approximate circular radius encompassing 90\% of \chandra's PSF (in arcsec). This radius was used to extract aperture photometry.
\newline \textit{Column 35}. H-band total counts within a circle of radius $r_{90}$.
\newline \textit{Column 36}. H-band background counts within a circle of radius $r_{90}$, extracted from our background maps at the position of the source.
\newline \textit{Column 37}. H-band net counts within a circle of radius $r_{90}$, or $3\sigma$ upper limit if the probability of being spurious of the source is higher than the threshold.
\newline \textit{Columns 38-39}. Positive and negative errors on H-band net counts, or $-99$ if the net counts are a $3\sigma$ upper limit.
\newline \textit{Column 40}. H-band total exposure in seconds at the position of the source.
\newline \textit{Column 41}. H-band net count rate in cts/s, or $3\sigma$ upper limit if the probability of being spurious of the source is higher than the threshold.
\newline \textit{Columns 42-43}. Positive and negative errors on H-band net count rate, or $-99$ if the net count rate is a $3\sigma$ upper limit.
\newline \textit{Column 44}. H-band flux, or $3\sigma$ upper limit if the probability of being spurious of the source is higher than the threshold. The flux is obtained from the count rate (or its upper limit) applying an aperture correction and the appropriate exposure-weighted energy conversion factor assuming $\Gamma=1.4$.
\newline \textit{Columns 45-46}. Positive and negative errors on the H-band flux, or $-99$ if the flux is a $3\sigma$ upper limit.
\newline \textit{Column 47}. HR computed with BEHR \citep{park06}.
\newline \textit{Columns 48-49}. Positive and negative errors on HR.
\newline \textit{Column 50}. Index of the XBOOTES counterpart in \citet{kenter05}, $-99$ if no counterpart is found.
\newline \textit{Column 51}. Name of the XBOOTES counterpart in \citet{kenter05}, $-99$ if no counterpart is found.
\newline \textit{Column 52}. Distance in arcsec between the CDWFS source and its XBOOTES counterpart, $-99$ if no counterpart is found.
\newline \textit{Column 53}. Number of \xb counterparts associated within $1.1 \times r_{90}$ to the CDWFS source. 0 if no counterpart is found. If more than one \xb counterpart is found, the closest one was chosen.
\newline \textit{Columns 54-55}. J2000 RA and DEC coordinates for the $I$-band NDWFS counterpart, $-99$ if no counterpart in $I$-band is found.
\newline \textit{Column 56}. Vega $I$-band automatic magnitude (``MAG\_AUTO") as computed by SExtractor \citep{bertin96}. $-99$ if no NDWFS counterpart in $I$-band is found, 99 if $I$-band photometry is unreliable (see also the ``FLAG\_DEEP" entry).
\newline \textit{Columns 57-58}. J2000 RA and DEC coordinates for the [3.6]-band SDWFS counterpart, $-99$ if no counterpart in [3.6]-band is found.
\newline \textit{Columns 59-66}. Vega SDWFS magnitudes for the four IRAC channels, and their uncertainties; $-99$ if no counterpart in [3.6]-band is found.
\newline \textit{Column 67}. SDWFS flag; $-99$ if no counterpart in [3.6]-band is found.
\newline \textit{Column 68}. Separation in arcsec between the CDWFS source and its NDWFS counterpart; NULL if no counterpart in $I$-band is found.
\newline \textit{Column 69}. Separation in arcsec between the CDWFS source and its SDWFS counterpart; NULL if no counterpart in [3.6]-band is found.
\newline \textit{Column 70}. Separation in arcsec between the NDWFS and SDWFS counterparts; NULL if no counterpart is found in either of the two bands.
\newline \textit{Column 71}. Maximal separation in arcsec between the CDWFS and its counterparts; 0.0 if no counterpart is found in both the two bands.
\newline \textit{Column 72}. Number of catalogs in which the source is contained, with 1 meaning only CDWFS and 3 meaning CDWFS, NDWFS, and SDWFS.
\newline \textit{Column 73}. NWAY logarithm of the ratio between prior and posterior from distance matching.
\newline \textit{Column 74}. NWAY distance posterior probability comparing the association against no association.
\newline \textit{Columns 75-76}. NWAY bias factors coming from the additional $I$-band and [3.6]-band magnitude information. If the magnitudes are unknown, the factors reduce to unity.
\newline \textit{Column 77}. Same as Column 74, after additional information is added from the magnitude of the counterparts.
\newline \textit{Column 78}. NWAY \texttt{p\_any} parameter; for each CDWFS entry, represents the probability that one of the association is the correct one. As stated in the NWAY manual, a low value of this parameter does not rule out the associated counterpart; it can simply mean that there is not enough information to declare it secure.
\newline \textit{Column 79}. Spectroscopic redshift from the AGES catalog, $-99$ if no spectroscopic redshift is available.
\newline \textit{Column 80}. Photometric redshift estimate from the Duncan et al. (2020, submitted) catalog, obtained through fitting of the Spectral Energy Distribution (SED) with a hybrid template $+$ machine learning technique; $-99$ if no counterpart is found. Specifically, we provide the median of the primary redshift peak above the 80\% highest probability density (HPD) credible interval (CI) for the photometric redshift posterior.
\newline \textit{Columns 81-82}. Minimum and maximum of the primary peak above the 80\% HPD CI (Column 80) from the Duncan et al. (2020, submitted) catalog, obtained through SED fitting with a hybrid template $+$ machine learning technique; $-99$ if no counterpart is found.
\newline \textit{Column 83}. ``FLAG\_DEEP" = 1 indicates that the NDWFS source has good data that can go to the full depth of the $B_wRI$ optical images - this flag excludes bad pixels, bleed trails, bright star halos and bright galaxies. $-99$ if no counterpart is found.
\newline \textit{Column 84}. Index of the source in the Duncan et al. (2020, submitted) catalog.
\newline \textit{Column 85}. Best redshift: spectroscopic redshift when available, photometric otherwise. $-99$ if no redshift is available.
\newline \textit{Column 86}. Redshift flag: 1 if the best redshift is spectroscopic, 0 if it is photometric, $-99$ if no redshift is available.
\newline \textit{Columns 87-89}. Logarithm of the column density estimated from fitting the HR-$z$ plane using the best redshift, and positive and negative uncertainties. Note that the uncertainties reflect the HR ones only, while the redshift is fixed. The minimum value is 20, while $-99$ if no column density was computed (due to a missing redshift).
\newline \textit{Columns 90-98}. Correction factors $k$ with their maximal and minimal values, for the three \chandra bands. The correction factor is defined as the ratio between the observed (obscured) flux and the unobscured one. The maximal and minimal values reflect the maximal and minimal column density estimates. Unity values mean that the source has log($N_{\rm H}/$cm$^{-2}$) $\leq 20$ and is considered unobscured. $-99$ if no redshift is available.
\newline \textit{Columns 99-101}. De-absorbed, rest frame luminosities in the F, S and H bands, obtained from the absorption-corrected fluxes and assuming $\Gamma=1.8$. $-99$ if no redshift is available, or if $z = 0$.
\newline \textit{Column 102}. De-absorbed, rest frame luminosity in the $2-10$ keV band, obtained from the absorption-corrected H-band luminosity and assuming $\Gamma=1.8$. $-99$ if no redshift is available, or if $z = 0$.

\section{Summary and Conclusions} \label{sec:summary}

In this paper, we presented a detailed overview of the CDWFS, a new ambitious \chandra survey in the \bootes field. The information about the release of the associated data products can be found in Appendix \ref{asec:products}. CDWFS is comprised of 281 \chandra observations in the \bootes field spanning 15 years of observations, for a total observing time of $3.4$ Ms. 
\par To analyze this large dataset in an homogeneous way, we built accurate simulations of the whole field, taking into account the change of the instrument over the years. The resulting catalog of \howmany X-ray point sources has a spurious fraction of sources of $\sim 1\%$ in each band. Taking into account this spurious fraction, together with source incompleteness and Eddington bias, accurate number counts were derived. This analysis confirmed the presence of an overdensity of hard X-ray sources in the corner of the \bootes\ field that has the deepest observations (i.e.\ the LaLa survey field), as previously noted by \citet{wang04}. The large number of X-ray sources detected corresponds to a resolved fraction of the CXB of \softcxb between $0.5$--$2$ keV and \hardcxb between $2$--$7$ keV.
\par The wealth of multiwavelength data in the \bootes\ field allowed us to assign redshifts to $\sim 94\%$ of our sources. We then used the hardness-ratio--redshift plane to derive estimates of obscuration, and intrinsic luminosities, for our sources. The unique and homogeneous coverage of the luminosity--redshift parameter space for CDWFS sources, together with the extensive multi-wavelength coverage of the \bootes\ field, makes the CDWFS a valuable addition to the \chandra X-ray legacy surveys. In particular, the catalog that we present and release in this paper is ideal for investigating AGN-galaxy co-evolution at the cosmic epoch where AGN activity and star formation are most concurrent.

\acknowledgments
We thank the anonymous referee for a thorough report that helped to strengthen the paper and make it clearer. \newline
This research has made (heavy) use of new data obtained by the \chandra \textit{X-ray Observatory} and data obtained from the \chandra Data Archive, as well as software provided by the \chandra X-ray Center (CXC).
\newline A.M. and R.C.H. acknowledge support by the NSF through grant numbers 1554584, and by NASA through grant numbers NNX15AP24G and \chandra GO award GO7-18130X. R.J.A. was supported by FONDECYT, grant number 1191124. S.C. acknowledges financial support from the Department of Science and Technology through the SERB Early Career Research grant and Presidency University through the Faculty Research and Professional Development Funds. The work of P.R.M.E. and D.S. was carried out at the Jet Propulsion Laboratory, California Institute of Technology, under a contract with NASA. K.J.D. acknowledges support from the ERC Advanced Investigator programme NewClusters 321271.

\facilities CXO, KPNO:2.1m, Mayall, MMT, Spitzer
\software {CIAO} \citep{fruscione06}, XSPEC \citep{arnaud96}, NWAY \citep{salvato18}, Astropy \citep{astropy13,astropy18}

\newpage
\bibliographystyle{aasjournal}
\bibliography{amasini}

\appendix
\renewcommand{\bottomfraction}{0.8}
\setcounter{bottomnumber}{2}
\renewcommand\thefigure{\thesection.\arabic{figure}}  
\renewcommand\thetable{\thesection.\arabic{table}}  
\setcounter{figure}{0}
\setcounter{table}{0}
\section{NDWFS, SDWFS and \textit{Gaia} astrometry} \label{asec:astrometry}

As explained in the main text, all 281 \chandra observations included in this paper have been re-aligned to a common astrometric reference. In particular, we pinned the new aspect solutions of all observations to the USNO-A2 astrometry, using the catalog of NDWFS optical counterparts to \xb sources \citep{brand06}. This procedure was chosen to simplify the matching before a careful source detection. However, given that the \textit{Gaia} mission is now the new astrometric reference, we made sure that all the observations and catalogues used in this work refer to the same astrometric system.
\par
\begin{figure*}
    \centering
    \includegraphics[width=0.48\textwidth]{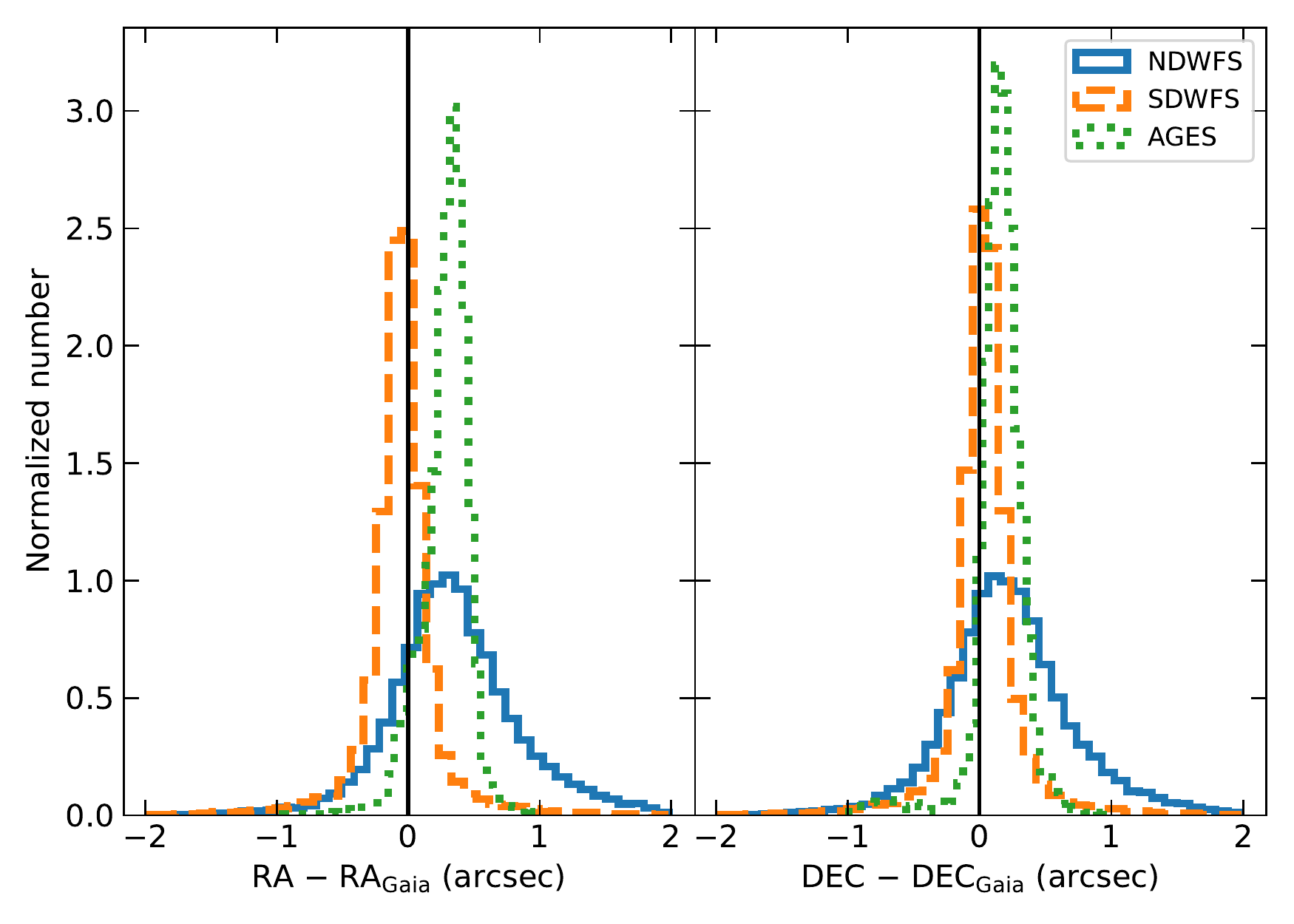}
    \includegraphics[width=0.48\textwidth]{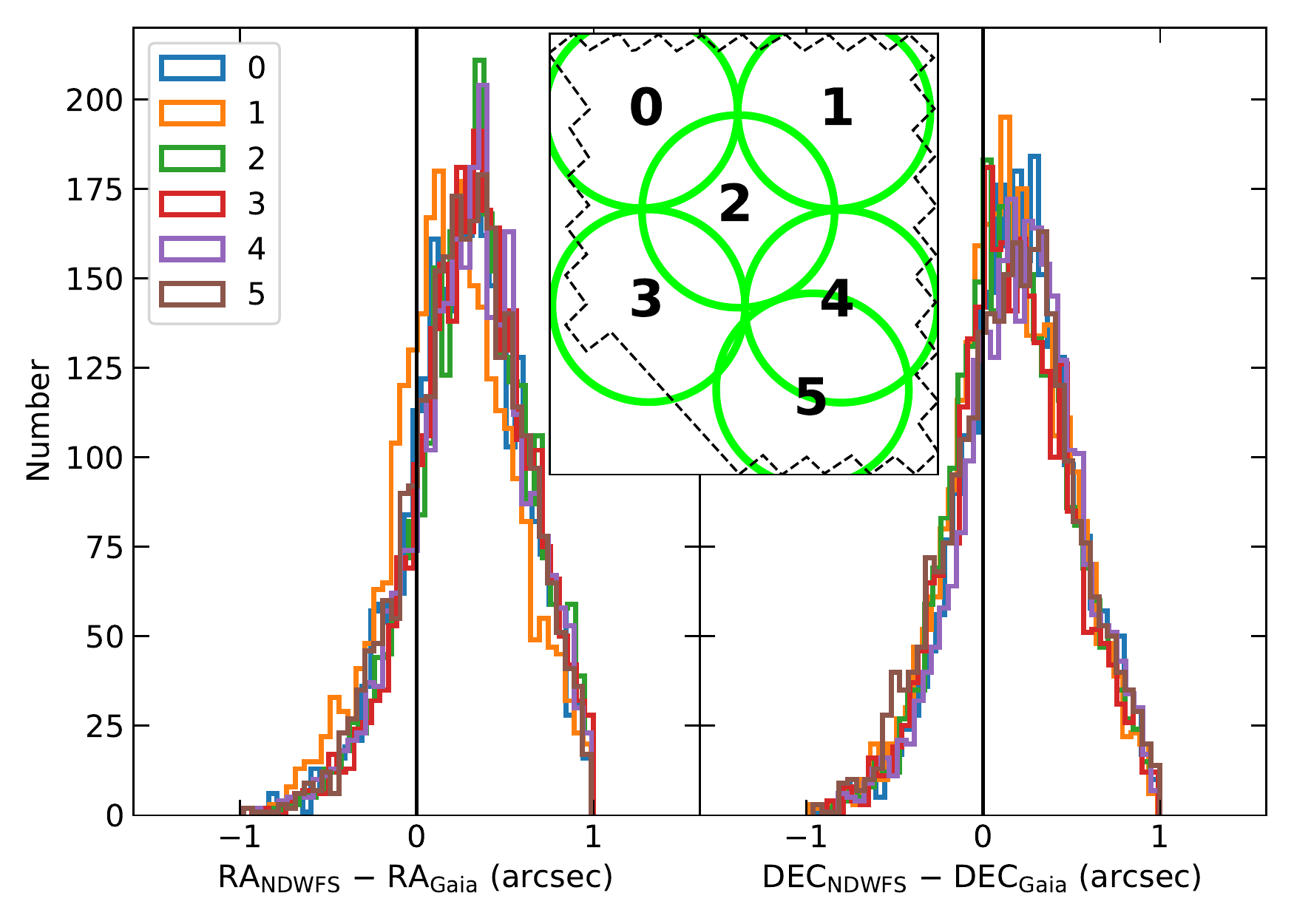}
    \caption{\textbf{Left.} NDWFS (solid blue histogram), AGES (dotted green histogram) and SDWFS (dashed orange histogram) astrometry compared to \textit{Gaia}. Unlike SDWFS, NDWFS and AGES show a systematic shift with respect to \textit{Gaia}. \textbf{Right.} NDWFS astrometric shift with respect to \textit{Gaia} for the six partially-overlapping regions highlighted in the inset. To be even more conservative about possible wrong associations, we matched sources within $1\arcsec$.}
    \label{afig:astrometry}
\end{figure*}

 To this aim, we cross matched the \textit{Gaia} DR2 catalog \citep{gaiadr2} to the USNO-A2 and the SDWFS catalogs, using a circle of 2 deg radius centered on RA = 218.0 deg, and DEC = +34.0 deg. As shown in the left panel of Figure \ref{afig:astrometry}, while the SDWFS coordinates did not show any systematic shift with respect to \textit{Gaia}, the NDWFS and AGES ones (both referring to the USNO-A2 WCS) appeared to be shifted by $\Delta$RA  $= 0\farcs32$ and $\Delta$DEC  $= 0\farcs20$. A similar shift was noticed between NDWFS and SDSS by \citet{cool07}. We verified that this systematic effect is consistent throughout the field, splitting up the \bootes field in six, partially overlapping circular regions of 0.8 deg radius. In Figure \ref{afig:astrometry} we show how the shifts compare among regions, demonstrating that all regions show the same systematic shift in DEC and all but one in RA (region 1 showing a slightly smaller offset in RA). \par Hence, we decided to shift all the CDWFS mosaics used for subsequent analysis by the same amount, defining the new CRVAL1 and CRVAL2 header keywords as 

\begin{equation}
    \begin{array}{ccc} \text{CRVAL1,new} = \text{CRVAL1,old} - 8.927566\times10^{-5} ~\text{degrees}, \\ \text{CRVAL2,new} = \text{CRVAL2,old} - 5.610275\times10^{-5} ~\text{degrees}. \end{array}
\end{equation}

\section{Tables of the excluded \xb sources and of the main catalog} \label{asec:xbootes}

Table \ref{atab:xb_excluded} shows an extract of the first ten rows of the additional list of \xb sources not included in the CDWFS catalog, while Table \ref{atab:main_catalog} gives an extract (with few selected columns) of the first ten rows of the main catalog released with this paper. Both full catalogs can be found in the electronic version of the paper. 

\begin{longrotatetable}
\begin{deluxetable}{cc cc cc cc cc cc cc cc cc c}
\tabletypesize{\scriptsize}
\tablecaption{Extract of the table with information on the excluded \xb sources. \label{atab:xb_excluded}}

\tablehead{\colhead{CXOXB} & \colhead{ID} & \colhead{RA} & \colhead{DEC} & \colhead{PROB\_F} & \colhead{R90\_F} & \colhead{TOT\_F} & \colhead{BKG\_F} & \colhead{EXP\_F} & \colhead{PROB\_S} & \colhead{R90\_S} & \colhead{TOT\_S} & \colhead{BKG\_S} & \colhead{EXP\_S} & \colhead{PROB\_H} & \colhead{R90\_H} & \colhead{TOT\_H} & \colhead{BKG\_H} & \colhead{EXP\_H} \\
\colhead{} & \colhead{} & \colhead{deg} & \colhead{deg} & \colhead{} & \colhead{arcsec} & \colhead{} & \colhead{} & \colhead{s} & \colhead{} & \colhead{arcsec} & \colhead{} & \colhead{} & \colhead{s} & \colhead{} & \colhead{arcsec} & \colhead{} & \colhead{} & \colhead{s}} 

\startdata
J142420.5+333922 & 1 & 216.08566 & 33.65609 & 0.0849 & 15.62 & 4.0 & 1.94 & 3243 & 0.3428 & 15.62 & 1.0 & 0.67 & 3097 & 0.1138 & 16.42 & 3.0 & 1.41 & 3202 \\
J142438.1+334245 & 14 & 216.15904 & 33.71254 & 0.0043 & 8.44 & 4.0 & 0.67 & 4124 & 0.0223 & 8.44 & 2.0 & 0.24 & 4061 & 0.0829 & 9.24 & 2.0 & 0.53 & 4097 \\
J142440.4+351921 & 18 & 216.16857 & 35.32244 & 0.0008 & 7.07 & 4.0 & 0.41 & 3903 & 0.0005 & 7.07 & 3.0 & 0.15 & 3889 & 0.2326 & 7.87 & 1.0 & 0.32 & 3863 \\
J142448.6+334540 & 43 & 216.20241 & 33.76126 & 0.0001 & 7.37 & 5.0 & 0.47 & 3624 & 0.0006 & 7.37 & 3.0 & 0.17 & 3583 & 0.0416 & 8.17 & 2.0 & 0.34 & 3653 \\
J142449.8+351815 & 50 & 216.20757 & 35.30437 & $4.1\times10^{-5}$ & 4.41 & 4.0 & 0.19 & 4387 & 0.0023 & 4.41 & 2.0 & 0.07 & 4382 & 0.0088 & 5.21 & 2.0 & 0.14 & 4346 \\
J142451.7+351411 & 56 & 216.21541 & 35.23646 & 0.0002 & 5.75 & 4.0 & 0.28 & 4143 & 0.0960 & 5.75 & 1.0 & 0.11 & 4106 & 0.0016 & 6.55 & 3.0 & 0.23 & 4091 \\
J142457.9+333318 & 67 & 216.24154 & 33.55514 & $3.5\times10^{-5}$ & 8.17 & 6.0 & 0.60 & 4056 & 0.1708 & 8.17 & 1.0 & 0.21 & 3935 & 0.0001 & 8.97 & 5.0 & 0.49 & 4072 \\
J142504.4+355015 & 83 & 216.26836 & 35.83769 & $7.6\times10^{-5}$ & 7.30 & 5.0 & 0.43 & 3980 & 0.0074 & 7.30 & 2.0 & 0.13 & 3985 & 0.0054 & 8.10 & 3.0 & 0.36 & 3950 \\
J142504.5+355125 & 84 & 216.26879 & 35.85704 & 0.0054 & 9.16 & 4.0 & 0.72 & 4331 & 0.0192 & 9.16 & 2.0 & 0.22 & 4307 & 0.0190 & 9.96 & 3.0 & 0.59 & 4332 \\ 
J142505.6+351303 & 89 & 216.27333 & 35.21770 & $4.1\times10^{-5}$ & 4.54 & 4.0 & 0.19 & 4344 & $5.3\times10^{-5}$ & 4.54 & 3.0 & 0.07 & 4344 & 0.1335 & 5.34 & 1.0 & 0.16 & 4330 \\
\enddata

\tablecomments{The F, S, and H subscripts refer to the F, S, and H bands, respectively. The ID column refers to the number of the source in the \citet{kenter05} catalog. The full version of the list can be found in the electronic version of the catalog.}

\end{deluxetable}
\end{longrotatetable}

\begin{longrotatetable}
\begin{deluxetable}{cc cc cc cc cc cc cc cc}
\tabletypesize{\scriptsize}
\tablecaption{Extract of the main source catalog table. \label{atab:main_catalog}}

\tablehead{\colhead{RAJ2000} & \colhead{DEJ2000} & \colhead{POS\_ERR} & \colhead{FLUX\_F} & \colhead{HR} & \colhead{XB\_ID} & \colhead{NDWFS\_RA} & \colhead{NDWFS\_DEC} & \colhead{I\_MAG\_AUTO} & \colhead{SDWFS\_RA} & \colhead{SDWFS\_DEC} & \colhead{SDWFS\_CH1\_MA} & \colhead{ZSP} & \colhead{ZPH} & \colhead{LOG\_NH} & \colhead{LINT\_F}\\
\colhead{deg} & \colhead{deg} & \colhead{arcsec} & \colhead{\fluxcgs} & \colhead{} & \colhead{} & \colhead{deg} & \colhead{deg} & \colhead{mag} & \colhead{deg} & \colhead{deg} & \colhead{mag} & \colhead{} & \colhead{} & \colhead{cm$^{-2}$} & \colhead{\lumcgs}} 

\startdata
218.09335 & 32.30161 & 2.12 & $2.70 \times 10^{-14}$ & $-0.808$ & 1999 & $-99$ & $-99$ & $-99$ & 218.09415 & 32.30075 & 15.17 & $-99$ & $-99$ & $-99$ & $-99$ \\
216.33986 & 32.31454 & 1.22 & $3.79 \times 10^{-14}$ & 0.065  & 170  & $-99$ & $-99$ & $-99$ & 216.33987 & 32.3142 & 18.31 & $-99$ & 3.1321 & 23.58 & $4.46 \times 10^{45}$ \\
217.67761 & 32.32548 & 0.80 & $1.17 \times 10^{-13}$ & $-0.608$ & 1552 & 217.67748  & 32.32556  & 18.45 & 217.67758 & 32.32557 & 15.40 & $-99$ & 0.4135 & 20.0  & $6.64 \times10^{43}$ \\
217.97970 & 32.34356 & 1.51 & $1.37 \times10^{-14}$ & 0.390  & 1872 & 217.97991 & 32.34344 & 23.06 & 217.97999 & 32.34352 & 17.34 & $-99$ & 1.8089 & 23.48 & $5.69 \times10^{44}$ \\
217.63251 & 32.36268  & 0.50 & $6.55\times10^{-14}$ & $-0.544$ & 1508 & 217.63243 & 32.36277 & 18.52 & 217.63245 & 32.36276 & 15.61 & $-99$ & 2.0131 & 21.54 & $1.60\times10^{45}$ \\
217.11544 & 32.37076 & 1.34 & $1.13\times10^{-14}$ & 0.908 & 953 & 217.11581 & 32.37097 & 23.73 & 217.11576 & 32.37104 & 17.96 & $-99$ & 1.899 & 24.06 & $1.03\times10^{45}$ \\
218.05901 & 32.37461 & 1.32 & $8.57\times10^{-15}$ & $-0.867$ & $-99$ & 218.05898 & 32.37503 & 23.94 & 218.05907 & 32.3751 & 16.97 & $-99$ & 1.6338 & 20.0 & $1.24\times10^{44}$ \\
216.71267 & 32.37187 & 1.0 & $1.59\times10^{-14}$ & 0.934 & 508 & 216.71255 & 32.37172 & 19.74 & 216.71231 & 32.3719 & 15.44 & $-99$ & 0.7764 & 23.48 & $1.46\times10^{44}$ \\
216.28777 & 32.37006 & 1.11 & $1.17\times10^{-14}$ & 0.643 & 105 & 216.28766 & 32.36974 & 23.51 & 216.28759 & 32.36986 & 18.47 & $-99$ & 2.0223 & 23.8 & $8.17\times10^{44}$ \\
217.72037 & 32.37844 & 1.66 & $1.25\times10^{-14}$ & $-0.162$ & 1589 & 217.72128 & 32.37761 & 25.15 & $-99$ & $-99$ & $-99$ & $-99$ & 1.0886 & 22.52 & $1.03\times10^{44}$ \\
\enddata

\tablecomments{The full electronic catalog, accessible through the online journal, contains many more columns, which are described in \S\ref{sec:catalog}.}

\end{deluxetable}
\end{longrotatetable}

\section{Public Data Products Associated with this Paper}\label{asec:products}
This paper publicly releases three catalogs:
\begin{enumerate}
\item The log of the 281 \chandra observations considered in this work (see Table \ref{tab:obs}).
\item The list of the \xbbelowthresh \xb sources not appearing in the main catalog (see Table \ref{atab:xb_excluded}).
\item The main catalog containing \howmany sources (see Table \ref{atab:main_catalog}).
\end{enumerate}
In addition, all the data products described in Section \ref{sec:data_red}, and the raw merged catalog returned by \texttt{wavdetect} (i.e., before reliability cuts) are available upon request.

\end{document}